\documentclass[1p]{elsarticle}
\biboptions{longnamesfirst,square,sort,compress}

\usepackage[english]{babel}
\usepackage[latin1]{inputenc}
\usepackage[T1]{fontenc}
\usepackage{graphicx}
\usepackage{enumerate}

\usepackage{amssymb,amsmath}
\usepackage{empheq}
\usepackage{mathtools}
\usepackage{mathrsfs} 
\usepackage{frcursive}
\usepackage{color}
\usepackage{comment}
\usepackage{multirow}
\usepackage{booktabs}
\usepackage{rotating}

\everymath{\displaystyle}
\usepackage{soul}
\allowdisplaybreaks[4]
\specialcomment{tomodify2}
{\begingroup\textcolor{red}}{\endgroup}
\specialcomment{myproofs}
{\begingroup\textcolor{blue}}{\endgroup}
\specialcomment{tothrow}
{\begingroup\textcolor{green}}{\endgroup}
\specialcomment{evol}
{\begingroup\textcolor{cyan}}{\endgroup}
\usepackage{hyperref}
\hypersetup{
    unicode=true,          
    pdftoolbar=true,        
    pdfmenubar=true,        
    pdffitwindow=false,     
    pdfstartview={FitH},    
    pdfnewwindow=true,      
    colorlinks=true,       
    linkcolor=red,          
    citecolor=green,        
    filecolor=magenta,      
    urlcolor=cyan           
}
\begin{document}

\begin{frontmatter}
	\title{Transverse shear warping functions for anisotropic multilayered plates}
	\author[drive]{A.~Loredo\corref{cor1}}
	\ead{alexandre.loredo@u-bourgogne.fr}
	%
	%
	\cortext[cor1]{Corresponding author}
	%
	\address[drive]{DRIVE, Université de Bourgogne, 49 rue Mlle Bourgeois, 58027 Nevers, France}
	%
%
%
%
\begin{abstract}
In this work, transverse shear warping functions for an equivalent single layer plate model are formulated from a variational approach. The part of the strain energy which involves the shear phenomenon is expressed in function of the warping functions and their derivatives. The variational calculus leads to a differential system of equations which warping functions must verify. Solving this system requires the choice of values for the (global) shear strains and their derivatives. A particular choice, which is justified for cross-ply laminates, leads to excellent results.
\par
For single layer isotropic and orthotropic plates, an analytical expression of the warping functions is given. They involve hyperbolic trigonometric functions. They differ from the $z-4/3z^3$ Reddy's formula which has been found to be a limit of present warping functions for isotropic and moderately thick plates. When the $h/L$ and/or the $G_{13}/E_{1}$ ratios significantly differ from those of isotropic and moderately thick plates, a difference between present warping functions and Reddy's formula can be observed, even for the isotropic single layer plate. Finite element simulations agree perfectly with the present warping functions in these cases.
\par
For multilayer cross-ply configurations, the warping functions are determined using a semi-analytical procedure. They have been compared to results of 3D finite element simulations. They are in excellent agreement. 
\par
For angle-ply laminates, the above process gives warping functions that seem to have relevant shapes, even if the choice for global shear values cannot be justified in this case. No finite element comparison has been presented at this time because it is difficult to propose boundary conditions and prescribed load that permit to isolate the shear phenomenon.
\end{abstract}
\end{frontmatter}
\section{Introduction}
The need to take into account the transverse shear behavior in plate theories has emerged when authors began to focus on thick homogeneous plates or on inhomogeneous plates. For homogeneous plates, the works of Reissner~\cite{Reissner1945}, Uflyand~\cite{Uflyand1948} and Mindlin~\cite{Mindlin1951} are considered as historical references. Industrial composite materials have appeared in the end of the 1930s. They can easily be used into multilayered structures, with some benefits. For these particular anisotropic materials, and also for multilayered plates, even for those constituted of isotropic plies, it has been rapidly evident that the weak and/or inhomogeneous transverse shear stiffness permit the shear deformation to significantly appear. Authors have then proposed models that fit this behavior. Among them, we can cite the early works of Lekhnitskii~\cite{Lekhnitskii1935} and Ambartsumyan~\cite{Ambartsumyan1958}. However, the one-layered models based on the Mindlin-Reissner model had been adapted to multilayered structures, giving the nowadays called \emph{first order shear deformation theory} (FOSDT). This theory is particularly efficient when shear correction factors are used. Those shear correction factors take into account the variations of the transverse shear stresses through the thickness. In what has become a reference work, Whitney~\cite{Whitney1973} gave a method for their computation. Other authors have later proposed different ways to compute them~\cite{Noor1989, Pai1995}.
\par
Since these works, static and dynamic behavior of anisotropic multilayered plates have been extensively studied, as it can be seen in Carrera's and Khandan's review articles~\cite{Carrera2002,Khandan2012}. Two important classes of plate theories that appear in such reviews are commonly called: the \emph{equivalent single layer theories} (ESL) and the \emph{layerwise theories} (LT). The latter contains theories that are based on hypothesis which are made in each layer, and lead to a number of variables which --generally-- increases with the number of layers. Some of them however propose an elimination process which lead to a number of variables which does not depend on the number of layers. On the other hand, the first class contains theories that are based on variables defined in a reference plane, and generally propose polynomial developments of fields over $z$. However, some of them, the \emph{higher order theories} propose advanced kinematic variations through the thickness, for example introducing warping functions which generally are built from the ply sequence. Accounting to these recent developments, it seems that there is a very thin line between these two class.    
\par
For all classes combined, the transverse shear strain variations through the thickness have recently received many attention, as it can be seen in references~\cite{Kim2006,Brischetto2009,Vidal2011,Neves2012,Dozio2012}. For the ESL class, the more recent and pertinent theories have been built using \emph{warping functions}~\cite{Kim2006,Loredo2013}. These \emph{warping functions} introduce the through-the-thickness ($z$) transverse shear strain variations directly in the layer displacement field. In these models, the first in-plane ($x$ and $y$) spatial derivatives of the (global) transverse shear strains are coupled with the membrane and bending terms, leading to a more complex stiffness matrix of size $10\times10$ compared to the classical $6\times6$ stiffness matrix of the FOSDT model. The behavior of these models are mainly determined by the choice of the warping functions. This is an interesting point because warping functions can be changed and/or adapted on the fly to fit particular requirement like special materials (adjacent plies with high Young's modulus ratio, viscoelastic behavior, functionally graded materials\dots) and/or in-plane inhomogeneous structures (damped structures with patches, structures with inserts, windows\dots). Indeed, the change of warping functions only requires a new computation of the stiffness and mass matrices of the model, which consists on evaluations of integrals over $z$. The process can therefore be completely automated. All the variables of the model remains the same, and keep the traditional meaning of ESL variables: in-plane displacements, transverse displacement, rotations and transverse shear, all these quantities being associated with a given reference plane.
\par
There are several techniques that can be used to formulate warping functions:
\begin{enumerate}
	\item One can try to built them by integration of the third equilibrium equation. Functions, which can be assimilated to warping functions, appear during the shear correction factor computation~\cite{Whitney1973}.
	\item One can also propose to search them into a reduced function family, like piecewise constant or quadratic functions of $z$, and then try to find the unknowns coefficients by means of kinematic and static conditions at interfaces~\cite{Pai1995}. An early work of Sun \& Whitney~\cite{Sun1973}, which has been later developed and enhanced to more general problems~\cite{Woodcock2008}, has been entirely reformulated in reference~\cite{Loredo2013} with the help of piecewise constant warping functions.
	\item They also can be built from an energetic variational approach~\cite{Kim2006}. In these approaches, it is also possible to restrain the functional space to constant or quadratic functions.  
\end{enumerate}
\par
In the present study, warping functions are obtained from an energetic variational approach. There is no \emph{a priori} hypothesis on their shapes. The main equations of the associated plate model~\cite{Loredo2013} are recalled here. Next, the part of the strain energy that involves the warping functions is isolated, and reformulated in order to make the variational calculus. It leads to a differential system of equations that warping functions must verify. The global values of the transverse shear strains and their spatial derivatives must be fixed in order to make the system solvable. A particular choice, which is justified for cross-ply laminates, gives excellent results. For angle-ply laminates, this choice is kept, but further work must be done to precise if this choice is relevant or if it exists a better way to do it. Then, an analytical solution is given, involving unknown constants for each layer of the multilayered structure. For a one-layer orthotropic plate, the constants are easy to determine and then the solution can be given in an analytical form. For multilayered plates, the solution requires the determination of these unknown constants which are related to continuity relations prescribed at the interfaces.
\section{Associated plate model}
In this section, the main equations of the associated plate model of reference~\cite{Loredo2013} is presented. In the following, Greek subscripts take values $1$ or $2$ and Latin subscripts take values $1$, $2$ or $3$. The Einstein's summation convention is used for subscripts only. The comma used as a subscript index means the partial derivative with respect to the directions corresponding to the following indexes.
\subsection{Laminate definition}
\begin{figure}
	\centering
		\includegraphics{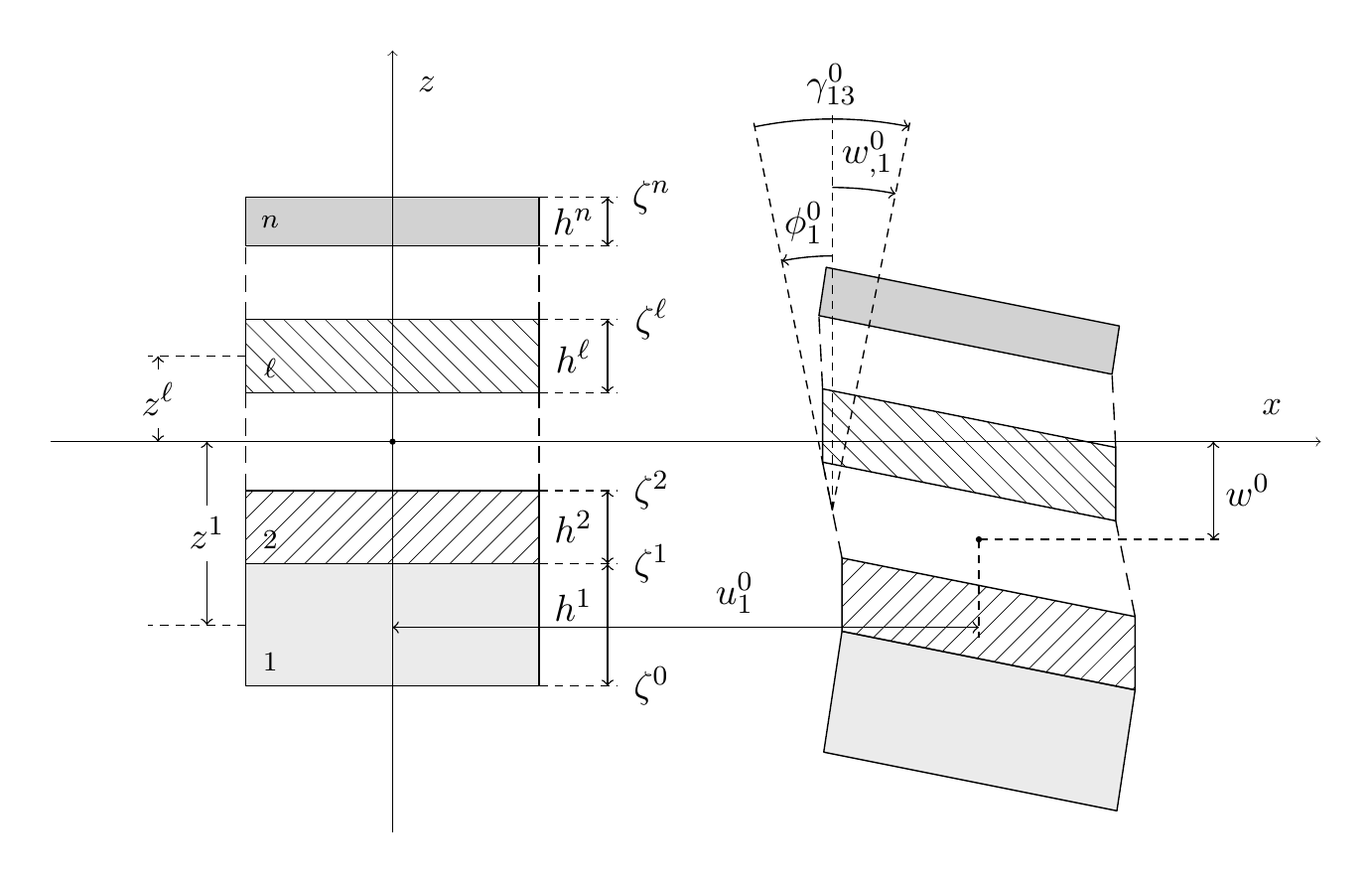}
	\caption{Geometrical parameters of the undeformed laminate on the left side, and its deformed shape with corresponding quantities on the right side.}
	\label{fig:Figure1}
\end{figure}

The laminate is composed of $n$ layers located between~$-h/2$ and $h/2$, where $h$ is the total height. The reference plane is taken as the $z=0$ plane. Figure~\ref{fig:Figure1} helps to visualize the following definitions:
\begin{itemize}[--] \topsep 0pt \itemsep 4pt \parskip 0pt
	\item $z^{\ell}$ is the elevation/offsetting of the middle plane of the layer $\ell$
	\item the $\ell$th layer is located between elevations $\zeta^{\ell-1}$ and $\zeta^{\ell}$
\end{itemize}
With these definitions:
\begin{itemize}[--] \topsep 0pt \itemsep 4pt \parskip 0pt
	\item there are $n$ parameters $z^{\ell}$,
	\item there are $n+1$ parameters $\zeta^{i}$ with $i$ taking values from $0$ to $n$,
	\item the thickness of the layer $\ell$ is $h^\ell=\zeta^{\ell}-\zeta^{\ell-1}$
\end{itemize}
\par
\subsection{Displacement field}\label{sec:DisplacementField}
The kinematic assumptions of the present formulation are:
\begin{equation}
\label{eq:depl_L}
  \left\{
    \begin{array}{lll}
      u_\alpha(x,y,z) & = & u_\alpha^0(x,y) - z w^0_{,\alpha}(x,y) + \varphi_{\alpha\beta}(z)\gamma^0_{\beta3}(x,y) \\ 
      u_3(x,y,z) & = & w^0(x,y) 
    \end{array} 
  \right.
\end{equation}
where $u_{\alpha}^0(x,y)$, $\gamma^0_{\alpha3}(x,y)$, and $w^0(x,y)$ are respectively the in-plane displacements, the engineering transverse shear strains\footnote{Note that if the reference plane coincides with an interface, this definition is not correct, but we shall see that it is always possible to define equivalent transverse strains.}, and the transverse displacement at the reference plane, and $\varphi_{\alpha\beta}(z)$ are four warping functions that will be determined.
\par
\subsection{Strain field}\label{sec:StrainField}
Let us compute the strain field from equations~\eqref{eq:depl_L}:
\begin{subequations}\label{eq:strains}
\begin{empheq}[left=\empheqlbrace]{align}
  \epsilon_{\alpha\beta}(x,y,z) & = \epsilon_{\alpha\beta}^0(x,y) - z w^0_{,\alpha\beta}(x,y) 
	       + \frac{1}{2}\left(\varphi_{\alpha\gamma}(z)\gamma^0_{\gamma3,\beta}(x,y)+\varphi_{\beta\gamma}(z)\gamma^0_{\gamma3,\alpha}(x,y)\right) \label{eq:in_plane_strains}\\ 
  \epsilon_{\alpha3}(x,y,z) & = \frac{1}{2}\varphi'_{\alpha\beta}(z)\gamma^0_{\beta3}(x,y) \label{eq:transverse_strains}\\ 
  \epsilon_{33}(x,y,z) & = 0 \phantom{\frac{1}{2}}
\end{empheq}
\end{subequations}
Note that $\epsilon_{\alpha\beta}^0$, $w^0_{,\alpha\beta}$, $\gamma^0_{\alpha3,\beta}$ and $\gamma^0_{\alpha3}$ form a set of $12$ independent generalized strains to consider in this model.
\par
\subsection{Stress field}
The strains obtained in formula~\eqref{eq:strains} permit to compute the stresses:
\begin{subequations}\label{eq:stresses}
\begin{empheq}[left=\empheqlbrace]{align}
  \sigma_{\alpha\beta}(x,y,z) & = Q_{\alpha\beta\gamma\delta}(z)\left(\epsilon_{\gamma\delta}^0(x,y) - z w^0_{,\gamma\delta}(x,y)
	       + \varphi_{\gamma\mu}(z)\gamma^0_{\mu3,\delta}(x,y)\right) \label{eq:in_plane_stresses}\\ 
  \sigma_{\alpha3}(x,y,z) & = C_{\alpha3\beta3}\varphi'_{\beta\mu}(z)\gamma^0_{\mu3}(x,y) \label{eq:transverse_stresses}\\ 
  \sigma_{33}(x,y,z) & = 0
\end{empheq}
\end{subequations} 
where $Q_{\alpha\beta\gamma\delta}$ are the reduced generalized plane strain stiffnesses. The vanishing of the term $\tfrac{1}{2}$ of equation~\eqref{eq:in_plane_strains} into equation~\eqref{eq:in_plane_stresses} is not evident. Let us demonstrate it, omitting the $x$, $y$, and $z$ coordinates:
\begin{align}
  \nonumber
	\frac{1}{2}Q_{\alpha\beta\gamma\delta}\left(\varphi_{\gamma\mu}\gamma^0_{\mu3,\delta}+\varphi_{\delta\mu}\gamma^0_{\mu3,\gamma}\right)
	&= \frac{1}{2}Q_{\alpha\beta\gamma\delta}\varphi_{\gamma\mu}\gamma^0_{\mu3,\delta}+\frac{1}{2}Q_{\alpha\beta\gamma\delta}\varphi_{\delta\mu}\gamma^0_{\mu3,\gamma} \\
	&= \frac{1}{2}Q_{\alpha\beta\gamma\delta}\varphi_{\gamma\mu}\gamma^0_{\mu3,\delta}+\frac{1}{2}Q_{\alpha\beta\delta\gamma}\varphi_{\gamma\mu}\gamma^0_{\mu3,\delta} \\
	\nonumber
	&= Q_{\alpha\beta\gamma\delta}\varphi_{\gamma\mu}\gamma^0_{\mu3,\delta}\phantom{\frac{1}{2}}
\end{align}
The vanishing of the term $\tfrac{1}{2}$ of equation~\eqref{eq:transverse_strains} into equation~\eqref{eq:transverse_stresses} is due to another reason:
\begin{align}
	\sigma_{\alpha3}=C_{\alpha3\beta3}\epsilon_{\beta3}+C_{\alpha33\beta}\epsilon_{3\beta}=2C_{\alpha3\beta3}\epsilon_{\beta3}
	                =C_{\alpha3\beta3}(z)\varphi'_{\beta\gamma}(z)\gamma^0_{\gamma3}(x,y)
\end{align}
\subsection{Kinematic and static assumptions of the model}
The model requires the continuity of in-plane displacements and transverse shear stresses at each of the $n-1$ interfaces:
\begin{subequations}
\begin{empheq}[left=\empheqlbrace]{align}
      \lim_{z \rightarrow \zeta^\ell-} u_\alpha(x,y,z) & = \lim_{z \rightarrow \zeta^\ell+} u_\alpha(x,y,z) \label{eq:cont_depl_L}\\ 
      \lim_{z \rightarrow \zeta^\ell-} \sigma_{\alpha3}(x,y,z) & = \lim_{z \rightarrow \zeta^\ell+}\sigma_{\alpha3}(x,y,z) \label{eq:cont_sigm_L}
\end{empheq}
\end{subequations}
for $\ell \in [1 \dots n-1]$. These conditions~\eqref{eq:cont_depl_L} and~\eqref{eq:cont_sigm_L} can now be written in terms of conditions on the warping functions:
\begin{subequations}
\begin{empheq}[left=\empheqlbrace]{align}
      &\lim_{z \rightarrow \zeta^\ell-} \varphi_{\alpha\beta}(z)  = \lim_{z \rightarrow \zeta^\ell+} \varphi_{\alpha\beta}(z) \label{eq:cont_depl_L2}\\ 
      &\lim_{z \rightarrow \zeta^\ell-} C_{\alpha3\beta3}(z)\varphi'_{\beta\gamma}(z)  = \lim_{z \rightarrow \zeta^\ell+} C_{\alpha3\beta3}(z)\varphi'_{\beta\gamma}(z) \label{eq:cont_sigm_L2}
\end{empheq}
\end{subequations}
Equations~\eqref{eq:cont_depl_L2} represent the continuity of the warping functions at the $n-1$ interfaces whereas equations~\eqref{eq:cont_sigm_L2} represent ``jump'' conditions on their derivatives.
\subsection{Strain energy}\label{sec:strain_energy}
It is possible to compute the strain energy density $\delta J=1/2\epsilon_{ij}\sigma_{ij}$ from formulas~\eqref{eq:strains} and~\eqref{eq:stresses} and integrate it over the thickness to obtain a strain energy surface density $J(x,y)$:
\begin{align}\label{eq:StrainEnergy1}
  J(x,y)
	 &=\frac{1}{2}\int_{-h/2}^{h/2}\epsilon_{ij}\sigma_{ij}\text{d}z
	  =\frac{1}{2}\int_{-h/2}^{h/2}\left(\epsilon_{\alpha\beta}\sigma_{\alpha\beta}+2\epsilon_{\alpha3}\sigma_{\alpha3}+\epsilon_{33}\sigma_{33}\right)\text{d}z \nonumber \\
	 &=\frac{1}{2}\int_{-h/2}^{h/2}
	  \left[\left(\epsilon_{\alpha\beta}^0 - z w^0_{,\alpha\beta} + \frac{1}{2}\left(\varphi_{\alpha\gamma}(z)\gamma^0_{\gamma3,\beta}
		  +\varphi_{\beta\gamma}(z)\gamma^0_{\gamma3,\alpha}\right)\right)\sigma_{\alpha\beta}
			+2\frac{1}{2}\varphi'_{\alpha\beta}(z)\gamma^0_{\beta3}\sigma_{\alpha3}\right]\text{d}z \nonumber \\
	 &=\frac{1}{2}\int_{-h/2}^{h/2}
	  \left[\left(\epsilon_{\alpha\beta}^0 - z w^0_{,\alpha\beta} + \varphi_{\alpha\gamma}(z)\gamma^0_{\gamma3,\beta}\right)\sigma_{\alpha\beta}
		  +\varphi'_{\alpha\beta}(z)\gamma^0_{\beta3}\sigma_{\alpha3}\right]\text{d}z
\end{align}
%
%
$x$ and $y$ have been omitted in the right hand side, for clarity. 
\subsection{Generalized forces}\label{sec:generalized_forces}
The strain energy can also be written
\begin{align}\label{eq:StrainEnergy2}
  J =\frac{1}{2}
	  \left[\epsilon_{\alpha\beta}^0 N_{\alpha\beta} - w^0_{,\alpha\beta} M_{\alpha\beta} + \gamma^0_{\gamma3,\beta} P_{\gamma\beta} + \gamma^0_{\beta3} Q_{\beta} \right]
\end{align}
naturally introducing the following quantities which are the generalized forces,
\begin{subequations}\label{eq:generalized_forces1}
\begin{empheq}[left=\empheqlbrace]{align}
  \{N_{\alpha\beta},M_{\alpha\beta},P_{\gamma\beta}\}&=\int_{-h/2}^{h/2} \{1,z,\varphi_{\alpha\gamma}(z)\} \sigma_{\alpha\beta}(z) \text{d}z \\
  Q_{\beta}&=\int_{-h/2}^{h/2} \varphi'_{\alpha\beta}(z) \sigma_{\alpha3}(z) \text{d}z
\end{empheq}
\end{subequations}
each associated with a corresponding generalized displacement in the strain energy formula~\eqref{eq:StrainEnergy2}. 
\par
$N_{\alpha\beta}$ and $M_{\alpha\beta}$ are respectively the classical plate membrane forces and bending moments, and $P_{\alpha\beta}$ and $Q_{\alpha}$ are special moments associated with the warping functions, \emph{i. e.} associated with the transverse shear behavior. Note that $P_{\alpha\beta} \ne P_{\beta\alpha}$ in the general case, leading to a set of $12$ generalized forces.
\subsection{Generalized stiffnesses}\label{sec:generalized_stiffnesses}
Let us calculate these generalized forces with the help of equations~\eqref{eq:stresses} and~\eqref{eq:generalized_forces1}. They are (some explanations are given below): 
\begin{subequations}\label{eq:generalized_forces2}
\begin{empheq}[left=\empheqlbrace]{align}
  N_{\alpha\beta} &=	A_{\alpha\beta\gamma\delta}\epsilon_{\gamma\delta}^0 + B_{\alpha\beta\gamma\delta} (-w^0_{,\gamma\delta})+ E_{\alpha\beta\mu\delta} \gamma^0_{\mu3,\delta}\\
	M_{\alpha\beta} &=	B_{\alpha\beta\gamma\delta}\epsilon_{\gamma\delta}^0 + D_{\alpha\beta\gamma\delta} (-w^0_{,\gamma\delta})+ F_{\alpha\beta\mu\delta} \gamma^0_{\mu3,\delta}\\
	P_{\alpha\beta} &=	E_{\gamma\delta\alpha\beta}\epsilon_{\gamma\delta}^0 + F_{\gamma\delta\alpha\beta} (-w^0_{,\gamma\delta})+ G_{\alpha\beta\mu\delta} \gamma^0_{\mu3,\delta}\\
	Q_{\alpha} &=	H_{\alpha3\beta3}\gamma^0_{\beta3}
\end{empheq}
\end{subequations}
where the following generalized stiffnesses have been introduced:
\begin{subequations}\label{eq:generalized stiffnesses1}
\begin{empheq}[left=\empheqlbrace]{align}
  \{A_{\alpha\beta\gamma\delta},B_{\alpha\beta\gamma\delta},D_{\alpha\beta\gamma\delta},E_{\alpha\beta\mu\delta},F_{\alpha\beta\mu\delta},G_{\nu\beta\mu\delta}\}&=\int^{\zeta^n}_{\zeta^0} Q_{\alpha\beta\gamma\delta}\{1,z,z^2,\varphi_{\gamma\mu}(z),z\varphi_{\gamma\mu}(z),\varphi_{\alpha\nu}(z)\varphi_{\gamma\mu}(z)\}\text{d}z \\
  H_{\alpha3\beta3}&=\int^{\zeta^n}_{\zeta^0} \varphi'_{\gamma\alpha}(z) C_{\gamma3\delta3} \varphi'_{\delta\beta}(z) \text{d}z
\end{empheq}
\end{subequations}
The $N_{\alpha\beta}$ and $M_{\alpha\beta}$ computation is straightforward but special attention must be paid to the calculus of the $P_{\alpha\beta}$, revealing some uncommon symmetries:
\begin{align}
	P_{\alpha\beta} 
	& =	\int^{\zeta^n}_{\zeta^0} \varphi_{\mu\alpha}(z) \sigma_{\mu\beta}(z) \text{d}z \nonumber \\
	& = \int^{\zeta^n}_{\zeta^0} \varphi_{\mu\alpha}(z) Q_{\mu\beta\gamma\delta}(z) \left(\epsilon_{\gamma\delta}^0 - z w^0_{,\gamma\delta} + \varphi_{\gamma\nu}(z)\gamma^0_{\nu3,\delta}\right) \text{d}z \nonumber \\
	& = E_{\gamma\delta\alpha\beta}\epsilon_{\gamma\delta}^0 + F_{\gamma\delta\alpha\beta} (-w^0_{,\gamma\delta})+ G_{\alpha\beta\mu\delta} \gamma^0_{\mu3,\delta} \phantom{\int^{\zeta^n}_{\zeta^0} }
\end{align}
In this last expression, the $E_{\gamma\delta\alpha\beta}$ and $F_{\gamma\delta\alpha\beta}$ are identified with the help of the major symmetry of the $Q_{\mu\beta\gamma\delta}(z)$ tensor.
\par 
The $\mathbf{A}$, $\mathbf{B}$ and $\mathbf{D}$ tensors inherit the symmetries of Hooke's tensor, a symmetry for each pair of indexes, called the \emph{minor symmetries} and the \emph{major symmetry} which permits the swap of the two pairs of indexes, this last one being related to the existence of a strain energy. The $\mathbf{E}$ and $\mathbf{F}$ tensors loose the symmetry on the last pair of indexes, forcing the major symmetry to disappear. The $\mathbf{G}$ tensor looses the symmetry on the two pairs of indexes, but keeps the major symmetry:
%
\begin{empheq}[left=\empheqlbrace]{alignat=2}\label{eq:tensors_symmetries}
  \nonumber 
	&\text{for the }\mathbf{A}\text{, }\mathbf{B}\text{, }\mathbf{D} \text{ tensors:} &\quad
	  &A_{\beta\alpha\gamma\delta}   =  A_{\alpha\beta\gamma\delta}   =  A_{\gamma\delta\alpha\beta}   =  A_{\gamma\delta\beta\alpha}  \\
  &\text{for the }\mathbf{E}\text{, }\mathbf{F}\text{ tensors:} &\quad
    &E_{\beta\alpha\gamma\delta}   =  E_{\alpha\beta\gamma\delta} \neq E_{\gamma\delta\alpha\beta} \neq E_{\gamma\delta\beta\alpha}  \\
  \nonumber 
	&\text{for the }\mathbf{G}\text{ tensor:} &\quad
    &G_{\beta\alpha\gamma\delta} \neq G_{\alpha\beta\gamma\delta}   =  G_{\gamma\delta\alpha\beta} \neq G_{\delta\gamma\beta\alpha} 
\end{empheq}
%
Hence there are $6$ independent components for $\mathbf{A}$, $\mathbf{B}$ and $\mathbf{D}$, $12$ for $\mathbf{E}$ and $\mathbf{F}$, $10$ for $\mathbf{G}$, and $3$ for $\mathbf{H}$. So this plate model has a total of $55$ independent stiffness coefficients in the most general case.
\subsection{Static laminate behavior}
This section aims to present the static\footnote{In reference~\protect\cite{Loredo2013}, the dynamic behavior is also developed, leading to generalized mass terms and motion equations.} laminate behavior in a matrix form.  The generalized forces are set, by type, into vectors:
\begin{equation}\label{eq:generalized_forces}
  \mathbf{N}=
  \begin{Bmatrix}
    N_{11} \\
		N_{22} \\
		N_{12}
  \end{Bmatrix}
  \quad
  \mathbf{M}=
  \begin{Bmatrix}
    M_{11} \\
		M_{22} \\
	  M_{12}
  \end{Bmatrix}
  \quad
  \mathbf{P}=
  \begin{Bmatrix}
    P_{11} \\
		P_{22} \\
		P_{12} \\
	  P_{21}
  \end{Bmatrix}
	\quad
  \mathbf{Q}=
  \begin{Bmatrix}
    Q_{1} \\
		Q_{2}
  \end{Bmatrix}
\end{equation} 
and the same is done for the corresponding generalized strains:
\begin{equation}\label{eq:generalized_strains}
  \boldsymbol{\epsilon}=
  \begin{Bmatrix}
    \epsilon^0_{11} \\
		\epsilon^0_{22} \\
		\epsilon^0_{12}
  \end{Bmatrix}
  \quad
  \boldsymbol{\kappa}=
  \begin{Bmatrix}
    -w^0_{,11} \\
		-w^0_{,22} \\
	  -2w^0_{,12}
  \end{Bmatrix}
  \quad
  \mathbf{\Gamma}=
  \begin{Bmatrix}
    \gamma^0_{13,1} \\
		\gamma^0_{23,2} \\
		\gamma^0_{13,2} \\
	  \gamma^0_{23,1}
  \end{Bmatrix}
  \quad
  \boldsymbol{\gamma}=
  \begin{Bmatrix}
    \gamma^0_{13} \\
		\gamma^0_{23}
  \end{Bmatrix}
\end{equation} 
Generalized forces are linked with the generalized strains by the $10\times10$ and $2\times2$ following behavior matrices:
\begin{equation}\label{eq:behavior}
\begin{Bmatrix}
  \mathbf{N} \\
	\mathbf{M} \\
	\mathbf{P}
\end{Bmatrix}
=
\begin{bmatrix}
  \mathbf{A} & \mathbf{B} & \mathbf{E} \\
	\mathbf{B} & \mathbf{D} & \mathbf{F} \\
	\mathbf{E^T} & \mathbf{F^T} & \mathbf{G}
\end{bmatrix}
\begin{Bmatrix}
  \boldsymbol{\epsilon} \\
	\boldsymbol{\kappa} \\
	\boldsymbol{\Gamma}
\end{Bmatrix}
\quad
\begin{Bmatrix}
  \mathbf{Q}
\end{Bmatrix}
=
\begin{bmatrix}
  \mathbf{H}
\end{bmatrix}
\begin{Bmatrix}
	\boldsymbol{\gamma}
\end{Bmatrix}
\end{equation} 
\section{Warping functions}
\subsection{Variational formulation}
Replacing the generalized forces in the strain energy of formula~\eqref{eq:StrainEnergy2} by means of formula~\eqref{eq:behavior}, and discarding all the quantities which do not depend on the shear phenomenon leads to:
\begin{align}\label{eq:StrainEnergy3}
  J_s =\frac{1}{2}
	  \left[
		\boldsymbol{\Gamma^T} 
		  \left(2\mathbf{E^T} \boldsymbol{\epsilon} + 2\mathbf{F^T} \boldsymbol{\kappa} +  \mathbf{G} \boldsymbol{\Gamma}\right) 
			+ 
			\boldsymbol{\gamma^T} \mathbf{H} \boldsymbol{\gamma}
		\right]
\end{align}
We shall develop in the following lines, a matrix form of stiffnesses definitions of formula~\eqref{eq:generalized stiffnesses1}. Introducing the following matrices:
\begin{equation}\label{eq:QCsAndPhiMatrices}
  \mathbf{C_m}
  = 
	\begin{bmatrix}
    Q_{1111} & Q_{1122} & Q_{1112} \\
    Q_{1122} & Q_{2222} & Q_{2212} \\
    Q_{1122} & Q_{2212} & Q_{1212} 
  \end{bmatrix}
  \quad , \quad
  \mathbf{C_s}
  = 
	\begin{bmatrix}
    C_{2323} & C_{2313}\\
    C_{2313} & C_{1313}
  \end{bmatrix}
  \quad , \quad
  \boldsymbol{\Phi}
	=
	\begin{bmatrix}
    \varphi_{11} &       0      &       0      & \varphi_{12} \\
          0      & \varphi_{22} & \varphi_{21} &       0      \\
    \varphi_{21} & \varphi_{12} & \varphi_{11} & \varphi_{22} 
  \end{bmatrix}
\end{equation} 
$z$ being omitted for clarity, one can verify that: 
\begin{equation}\label{eq:EFGH}
	\mathbf{E} = \int_{-h/2}^{h/2} \mathbf{C_m} \boldsymbol{\Phi} \text{d}z	
  \quad , \quad
  \mathbf{F} = \int_{-h/2}^{h/2} \mathbf{C_m} \boldsymbol{\Phi} z \text{d}z	
  \quad , \quad
  \mathbf{G} = \int_{-h/2}^{h/2} \boldsymbol{\Phi}\mathbf{^T} \mathbf{C_m} \boldsymbol{\Phi} \text{d}z	
  \quad , \quad
  \mathbf{H} = \int_{-h/2}^{h/2} \boldsymbol{\Phi'}\mathbf{^T} \mathbf{C_s} \boldsymbol{\Phi'} \text{d}z
\end{equation}
Now, the shear contribution to strain energy can be written:
\begin{align}\label{eq:StrainEnergy4}
  J_s =\frac{1}{2}
	  \int_{-h/2}^{h/2}
		\left[
		\boldsymbol{\Gamma^T} 
		  \left(
			  2 \boldsymbol{\Phi}\mathbf{^T} \mathbf{C_m} \boldsymbol{\epsilon} 
				+ 2 z \boldsymbol{\Phi}\mathbf{^T} \mathbf{C_m} \boldsymbol{\kappa} 
				+ \boldsymbol{\Phi}\mathbf{^T} \mathbf{C_m}  \boldsymbol{\Phi}\boldsymbol{\Gamma}
			\right) 
			+ \boldsymbol{\gamma^T} \boldsymbol{\Phi'}\mathbf{^T} \mathbf{C_s} \boldsymbol{\Phi'} \boldsymbol{\gamma}
		\right] \text{d}z
\end{align}
\par 
For computational reasons, it is easier to search for continuous functions. This is the reason why we shall consider in the following, the ``stress warping functions''
\begin{equation}
  \psi_{\alpha\beta}(z) = C_{\alpha3\beta3}(z)\varphi_{\beta\gamma}(z)
\end{equation}
which are four functions of class $C^1$. For the following calculations, we need to arrange these four functions and their derivative in $4 \times 1$ vectors:
\begin{equation}\label{eq:PsiVector}
   \boldsymbol{\Psi}
	=
	\begin{Bmatrix}
    \psi_{11} \\
    \psi_{21} \\
    \psi_{22} \\
    \psi_{12} 
  \end{Bmatrix}
	\quad \text{and} \quad
   \boldsymbol{\Psi'}
	=
	\begin{Bmatrix}
    \psi'_{11} \\
    \psi'_{21} \\
    \psi'_{22} \\
    \psi'_{12} 
  \end{Bmatrix}	
\end{equation} 

Let us consider the matrices:
\begin{equation}\label{eq:MatrixZm}
  \mathbf{Z_m}
  =
	4
	\begin{bmatrix}
    S_{1313} \gamma^0_{13,1} & S_{1323}  \gamma^0_{13,1} & S_{1323}  \gamma^0_{23,1}  & S_{1313}  \gamma^0_{23,1} \\
		S_{2313} \gamma^0_{13,2} & S_{2323} \gamma^0_{13,2} & S_{2323} \gamma^0_{23,2} & S_{2313} \gamma^0_{23,2} \\
		S_{1313} \gamma^0_{13,2} + S_{2313} \gamma^0_{13,1} & S_{2323} \gamma^0_{13,1} + S_{1323} \gamma^0_{13,2} & S_{1323} \gamma^0_{23,2} + S_{2323} \gamma^0_{23,1} & S_{2313} \gamma^0_{23,1} + S_{1313} \gamma^0_{23,2} 
  \end{bmatrix}
\end{equation}
and:
\begin{equation}\label{eq:MatrixZs}
  \mathbf{Z_s}
  =
	4
	\begin{bmatrix}
    S_{1313} \gamma^0_{13} & S_{1323} \gamma^0_{13} & S_{1323} \gamma^0_{23} & S_{1313} \gamma^0_{23} \\
		S_{2313} \gamma^0_{13} & S_{2323} \gamma^0_{13} & S_{2323} \gamma^0_{23} & S_{2313} \gamma^0_{23} 
  \end{bmatrix}
\end{equation} 
It can be shown that: 
\begin{align}\label{eq:ChangeVars}
  \mathbf{Z_m} \boldsymbol{\Psi} = \boldsymbol{\Phi}\boldsymbol{\Gamma}
	\quad \text{and} \quad
	\mathbf{Z_s} \boldsymbol{\Psi'} = \boldsymbol{\Phi'} \boldsymbol{\gamma}
\end{align}
Now it is possible to reformulate the shear contribution to strain energy in a manner that the warping functions are the variables of a quadratic form:
\begin{align}\label{eq:StrainEnergy5}
  J_s =\frac{1}{2}
	  \int_{-h/2}^{h/2}
		\left[
		\boldsymbol{\Psi}\mathbf{^T} 
		  \left(
			  2 \mathbf{Z_m^T} \mathbf{C_m} \boldsymbol{\epsilon} 
				+ 2 z \mathbf{Z_m^T} \mathbf{C_m} \boldsymbol{\kappa} 
				+ \mathbf{Z_m^T} \mathbf{C_m} \mathbf{Z_m} \boldsymbol{\Psi}
			\right) 
			+  \boldsymbol{\Psi'}\mathbf{^T} \mathbf{Z_s^T} \mathbf{C_s} \mathbf{Z_s} \boldsymbol{\Psi'}
		\right] \text{d}z
\end{align}
The variationnal calculs leads to:
\begin{align}\label{eq:DeltaStrainEnergy1}
  \delta J_s =
	  \int_{-h/2}^{h/2}
		\left[
  	\delta\boldsymbol{\Psi}\mathbf{^T} 
		  \left(
			       \mathbf{Z_m^T} \mathbf{C_m} \boldsymbol{\epsilon} 
				+  z \mathbf{Z_m^T} \mathbf{C_m} \boldsymbol{\kappa} 
				+    \mathbf{Z_m^T} \mathbf{C_m} \mathbf{Z_m} \boldsymbol{\Psi}
			\right) 
			+  \delta\boldsymbol{\Psi'}\mathbf{^T} \mathbf{Z_s^T} \mathbf{C_s} \mathbf{Z_s} \boldsymbol{\Psi'}
		\right] \text{d}z
\end{align}
Note that two factor $2$ had appeared during the derivation of the third and fourth terms, and then vanished with all others during the simplification. Integrating by parts the fourth term leads to:
\begin{align}\label{eq:DeltaStrainEnergy2}
  \delta J_s =
	  \int_{-h/2}^{h/2}
		\left[
  	\delta\boldsymbol{\Psi}\mathbf{^T} 
		  \left(
			       \mathbf{Z_m^T} \mathbf{C_m} \boldsymbol{\epsilon} 
				+  z \mathbf{Z_m^T} \mathbf{C_m} \boldsymbol{\kappa} 
				+    \mathbf{Z_m^T} \mathbf{C_m} \mathbf{Z_m} \boldsymbol{\Psi}
			\right) 
			-  \delta\boldsymbol{\Psi}\mathbf{^T} \mathbf{Z_s^T} \mathbf{C_s} \mathbf{Z_s} \boldsymbol{\Psi''}
		\right] \text{d}z
			+  \left[\delta\boldsymbol{\Psi}\mathbf{^T} \mathbf{Z_s^T} \mathbf{C_s} \mathbf{Z_s} \boldsymbol{\Psi'}\right]^{h/2}_{-h/2}
\end{align} 
The last term is null if there is no prescribed shear stresses at the top and the bottom of the laminate. For concision, we shall denote:
\begin{align}\label{eq:Notations}	 
	\mathbf{H_z} = \mathbf{Z_s^T} \mathbf{C_s} \mathbf{Z_s}
	\:;\:
	\mathbf{G_z} = \mathbf{Z_m^T} \mathbf{C_m} \mathbf{Z_m}
	\:;\:
	\mathbf{F_z} = z \mathbf{Z_m^T} \mathbf{C_m}
	\:;\:
	\mathbf{E_z} = \mathbf{Z_m^T} \mathbf{C_m}
\end{align} 
 hence the differential system of equations that must verify the $\boldsymbol{\Psi}$ functions is written:
\begin{align}\label{eq:FinalSystem1}	 
	\mathbf{H_z} \boldsymbol{\Psi''} 
	=
	    \mathbf{G_z} \boldsymbol{\Psi}
	+   \mathbf{F_z} \boldsymbol{\kappa} 
	+   \mathbf{E_z} \boldsymbol{\epsilon} 
\end{align} 
\subsection{Solving the system}\label{sec:SolvingSystem}
If the matrices $\mathbf{H_z}$ and $\mathbf{G_z}$ have been invertible, the system~\eqref{eq:FinalSystem1} would have admit a general solution of the form
\begin{align}\label{eq:FinalSystem2}	 
 \boldsymbol{\Psi} 
	=	\mathbf{C_1} e^{(\mathbf{H_z}^{-1}\mathbf{G_z})^{1/2}z} + \mathbf{C_2} e^{-(\mathbf{H_z}^{-1}\mathbf{G_z})^{1/2}z} - \mathbf{G_z}^{-1} \mathbf{F_z} \boldsymbol{\kappa} - \mathbf{G_z}^{-1} \mathbf{E_z} \boldsymbol{\epsilon} 
\end{align} 
where $\mathbf{C_1}$ and $\mathbf{C_2}$ are two $1\times4$ vectors of constants defined for each layer, which must be adapted to make the warping functions and their derivatives verify conditions on top and bottom of the laminate, at the reference plane, and at the interfaces (detail is given below). Unfortunately, for the most general case, $\mathbf{H_z}$ is of rank $2$, $\mathbf{G_z}$ is of rank $3$ and they are both of dimensions $4 \times 4$. One can try to use generalized inverses, but the calculation is complicated and has not be found to give good results. The matrix $\mathbf{H_z}$ is written:
\begin{equation}\label{eq:MatrixHz}
  \mathbf{H_z}
  = 
	4
	\begin{bmatrix}
    S_{1313}      (\gamma^0_{13})^2     & S_{2313}      (\gamma^0_{13})^2     & S_{2313} \gamma^0_{13}\gamma^0_{23} & S_{1313} \gamma^0_{13}\gamma^0_{23} \\
		S_{1323}      (\gamma^0_{13})^2     & S_{2323}      (\gamma^0_{13})^2     & S_{2323} \gamma^0_{13}\gamma^0_{23} & S_{1323} \gamma^0_{13}\gamma^0_{23} \\
		S_{1323} \gamma^0_{13}\gamma^0_{23} & S_{2323} \gamma^0_{13}\gamma^0_{23} & S_{2323}      (\gamma^0_{23})^2     & S_{1323}      (\gamma^0_{23})^2     \\ 
		S_{1313} \gamma^0_{13}\gamma^0_{23} & S_{2313} \gamma^0_{13}\gamma^0_{23} & S_{2313}      (\gamma^0_{23})^2     & S_{1313}      (\gamma^0_{23})^2     
  \end{bmatrix}
\end{equation}
The general solution depends on values of $\gamma^0_{13}$ and $\gamma^0_{23}$, on values of their derivatives $\gamma^0_{13,1}$, $\gamma^0_{23,2}$, $\gamma^0_{13,2}$, $\gamma^0_{23,1}$ present in the matrix $\mathbf{G_z}$ and also on values of $\kappa^0_{11}$, $\kappa^0_{22}$, $\kappa^0_{12}$, $\epsilon^0_{11}$, $\epsilon^0_{22}$, $\epsilon^0_{12}$.
\par
It is possible to compute locally these values, starting with a classical plate model, and then to compute the warping functions to build the refined model, which will give new values for these quantities, and so on, with no guarantee of convergence. 
\par
Another approach consists on giving chosen values of the shear strains and their derivatives. For example, the choice $\gamma^0_{13}=1$, $\gamma^0_{23}=0$, $\gamma^0_{13,1}=-\pi/L_x$, $\gamma^0_{23,2}=0$, $\gamma^0_{13,2}=0$, $\gamma^0_{23,1}=0$ will give a reduced system of size $2\times 2$, only involving regular matrices, from which we can compute $\psi_{11}$ and $\psi_{21}$. The choice of $-\pi/L_x$ for the $\gamma^0_{13,1}$ value will be discussed later. The corresponding choice for the $y$ direction will give $\psi_{22}$ and $\psi_{12}$.
\subsection{Analytical solution for a single layer isotropic plate}
For an isotropic homogeneous plate, the system~\eqref{eq:FinalSystem1} reduces to two similar scalar differential equations whose solution is:
\begin{equation}\label{eq:OneLayerIsotropicSolution}
   \psi_{11}(z)=\frac{G}{a_1-1} \left( a_1 z -\frac{1}{a_2} \sinh \left( a_2 z \right)  \right)
 \end{equation}
where:
\begin{equation}
    a_1 = \cosh \left( a_2 \frac{h}{2} \right) \qquad \text{;} \qquad a_2=\frac{\gamma^0_{13,1}}{\gamma^0_{13}} \sqrt{\frac{Q}{G}}
\end{equation}
The other equation gives $\psi_{22}(z)$ which can be obtained replacing $\gamma^0_{13,1}$ by $\gamma^0_{23,2}$ and $\gamma^0_{13}$ by $\gamma^0_{23}$ in the above formulas. 
\par
The corresponding (strain) warping functions $\varphi_{\alpha\beta}$ are obtained dividing the above function by $G$. In addition, it is possible to reduce the warping functions in a manner that they take values only between $-1/2$ and $1/2$ applying the rule $\varphi^n_{11}(z)= \tfrac{1}{h} \varphi_{11}(h z)$. With this choice, we obtain:
\begin{equation}
   \varphi^n_{11}(z)=\frac{1}{a_1-1} \left( a_1 z  -\frac{1}{a_2 h } \sinh \left( a_2 h z \right)  \right)
\end{equation}
The choice of the values of $\gamma^0_{\alpha3,\beta}$ and $\gamma^0_{\alpha3}$ is done considering the Navier's or Levy's procedures. In these well known bending problems, the plate quantities (pressure, transverse displacements, transverse shears, etc.) are taken as trigonometric functions. For example for a rectangular plate with $0\le x \le L_x$ and $0\le y \le L_y$ the choice lead to $\gamma_{13}(x,y)=\gamma_{13}^{\text{max}}\cos(\pi x /L_x)\sin(\pi y /L_y)$. The derivative of this function with respect to $x$ can be written $\gamma_{13,1}(x,y)=\gamma_{13,1}^{\text{max}}\sin(\pi x /L_x)\sin(\pi y /L_y)$ that permits to write $\gamma_{13,1}^{\text{max}}=-\pi/L_x\gamma_{13}^{\text{max}}$. It is the reason why we choose the values $\gamma^0_{13}=1$, $\gamma^0_{13,1}=-\pi/L_x$. In addition, for an isotropic material, $Q/G=2/(1-nu)$, so one can calculate $a_1$ and $a_2$ to obtain the solution which only depends on $\nu$ and $h/L_x$:
\begin{equation}
   \varphi^n_{11}(z)=\frac{\left( \cosh\left(\frac{\pi h}{2L_x}\sqrt{\frac{2}{1-\nu}}\right) z  -\frac{L_x}{ h }\sqrt{\frac{1-\nu}{2}} \sinh \left( - \frac{\pi h}{L_x}\sqrt{\frac{2}{1-\nu}}  z \right)  \right)}{\cosh\left(\frac{\pi  h}{2L_x}\sqrt{\frac{2}{1-\nu}}\right)-1} 
\end{equation}  
It is interesting to compute a series expansion of this function:
\begin{equation}
   \varphi^n_{11}(z)= z + d_3 z^3 + d_5 z^5 + O(z^6)
\end{equation}
Table~\ref{tab:Coefficients} gives values of these coefficients for $\nu=0.3$ and $h/L_x \in \{0.0001, 0.001, 0.01, 0.1, 0.2, 0.4, 0.8\}$. It permits to evaluate the difference between this function and the Reddy's formula $z-4/3 z^3$. Some of the corresponding warping function are also plotted and compared to finite elements simulations in section~\ref{sec:Results}, figure~\ref{fig:IsotropeElancement}. 
\begin{table*}[htbp]
	\centering
		\begin{tabular}{cccccccc}
			 $h/L$ & 0.0001 & 0.001 & 0.01 & 0.1 & 0.2 & 0.4 & 0.8 \\
			\hline
			 $d_3$ &  -1.3333332      &  -1.3333325      &  -1.33326        & -1.326     & -1.302     & -1.215  & -0.9277 \\ 
			 $d_5$ & $-1.880 10^{-8}$ & $-1.880 10^{-6}$ & $-1.880 10^{-4}$ & -0.01869   & -0.07345   & -0.2740 & -0.8371 
		\end{tabular}
	\caption{Coefficients $d_3$ and $d_5$ of the series expansion of $\varphi^n_{11}(z)$ for the isotropic material.}
	\label{tab:Coefficients}
\end{table*}
\subsection{Analytical solution for a single layer orthotropic plate}
The orthotropic single layer case is interesting because, as we shall see later, it permit to set arbitrary small values for the transverse shear modulus. Warping functions that differ strongly from Reddy's formula can appear in these cases. There is another interest: the orthotropic layer can be put in an off-axis configuration, that lead to non null functions $\varphi_{12}$ and $\varphi_{21}$. We shall see that, due to the tensorial nature of the $\varphi_{\alpha\beta}$, it is possible to give exact formulas for the warping functions of an off-axis single layer orthotropic laminate. 
\par
For an on-axis orthotropic homogeneous plate, the system~\eqref{eq:FinalSystem1} reduces to two uncoupled $2\times2$ matrix differential equations. In addition, the matrix $\mathbf{H_z}^{-1}\mathbf{G_z}$ is diagonal. When the interface and top/bottom/middle conditions are enforced (see explanations in section~\ref{sec:MultiCrossPlyOrtho}), the $\psi_{21}(z)$ (for the first system) and the $\psi_{12}(z)$ (for the second system) are found to be null functions. Hence, the solution is:
\begin{empheq}[left=\empheqlbrace]{align}\label{eq:OneLayerOrthotropicSolution}
   \psi_{11}(z)&=\frac{G_{13}}{a_1-1} \left( a_1 z -\frac{1}{a_2} \sinh \left( a_2 z \right)  \right) \\
	 \psi_{22}(z)&=\frac{G_{23}}{a_3-1} \left( a_3 z -\frac{1}{a_4} \sinh \left( a_4 z \right)  \right)
 \end{empheq}
where:
\begin{empheq}[left=\empheqlbrace]{align}
    a_1 &= \cosh \left( a_2 \frac{h}{2} \right) \qquad \text{;} \qquad a_2=\frac{\gamma^0_{13,1}}{\gamma^0_{13}} \sqrt{\frac{Q_{1111}}{G_{13}}} \\
    a_3 &= \cosh \left( a_4 \frac{h}{2} \right) \qquad \text{;} \qquad a_4=\frac{\gamma^0_{23,2}}{\gamma^0_{23}} \sqrt{\frac{Q_{2222}}{G_{23}}}
\end{empheq}
\par
As for the isotropic case, the warping functions $\varphi_{11}(z)$ and $\varphi_{22}(z)$ are obtained from $\psi_{11}(z)$ and $\psi_{22}(z)$ by division by $G_{13}$ and $G_{23}$ respectively. It is also possible to have reduced warping functions with $z$ varying from $-1/2$ to $1/2$, in order to compare them for different ratios $h/L$. The choice for ratios values of $\gamma^0_{13,1}/\gamma^0_{13}=-\pi/L_x$ and $\gamma^0_{23,2}/\gamma^0_{23}=-\pi/L_y$ is also applied. Although it is not done in this paper, it is then possible to study rectangular plates. As we shall see later, three studies are done for this case, one with various ratios $h/L$, another with various ratios $G_{13}/E_1$ (we focus on the $x$ direction), and an off-axis study. 
\par
For this last case, due to their tensor nature, warping functions $\varphi_{\alpha\beta}$ must obey to classical coordinate transformation formulas. Hence, the warping functions of a $\theta$-oriented single ply can be calculated from the warping functions of the on-axis one:
\begin{empheq}[left=\empheqlbrace]{align}\label{eq:phitheta}
  \varphi^{\theta}_{11}&=c^2\varphi^0_{11}+s^2\varphi^0_{22} \\
  \varphi^{\theta}_{22}&=s^2\varphi^0_{11}+c^2\varphi^0_{22} \\
  \varphi^{\theta}_{12}&=-cs\varphi^0_{11}+sc\varphi^0_{22} \\
  \varphi^{\theta}_{21}&=-cs\varphi^0_{11}+sc\varphi^0_{22}
\end{empheq}
where $c=\cos(\theta)$ and $s=\sin(\theta)$. Note that if the material have a square symmetry and if the height to length ratios are the same in each direction, $\varphi^0_{11}=\varphi^0_{22}$ and then the $\varphi^{\theta}_{12}$ and $\varphi^{\theta}_{21}$ functions are null, but if it is not the case, they are not null. However, the generic algorithm, described in the following section, gives functions $\varphi^{\theta}_{12}$ and $\varphi^{\theta}_{21}$ that are similar in shape to those of formulas~\eqref{eq:phitheta} but do not present the symmetry which is evident in the above formulas. This must be studied in the future. The functions given by the code are shown in figure~\ref{fig:OrthotropicSingleLayer30}.   
\subsection{Multilayered cross-ply orthotropic plates}\label{sec:MultiCrossPlyOrtho}
As it was told before, for on-axis orthotropic plies, the system~\eqref{eq:FinalSystem1} reduces to two uncoupled $2\times2$ matrix differential equations. Hence the system is split into two subsystems for which the solution~\eqref{eq:FinalSystem2} can be computed. The first subsystem consists of the first two lines of system~\eqref{eq:FinalSystem1}, it permits to compute $\psi_{11}$ and $\psi_{21}$. 
\par
There are two unknown constants per layer in $\mathbf{C_1}$ and two in $\mathbf{C_2}$. There are also three global constants in $\boldsymbol{\kappa}$ and three in $\boldsymbol{\epsilon}$. The warping functions must verify 2 conditions at the reference plane $\varphi_{\alpha1}(0)=0$. The derivatives of the warping functions must verify 4 conditions at the top and the bottom of the plate $\varphi'_{\alpha1}(\pm h/2)=0$ and 2 conditions at the reference plane $\varphi'_{\alpha1}(0)=\delta_{\alpha1}$. These conditions can be easily transformed into conditions for the $\psi$ stress warping functions. Hence, for a single layer plate, there are $2+2+3+3=10$ constants to determine with $4+2+2=8$ conditions. If the unknowns are determined using a least square solving method, the $\kappa_{12}$ and $\varepsilon_{12}$ constant are found to be null, which is coherent with the on-axis orthotropic behavior. Hence we have the choice to eliminate them and solve a square system, or to keep them and use a least square method. 
\par
For multilayered plates, the warping functions $\varphi_{\alpha1}$ must be continuous at each interface, which gives 2 conditions per interface, and the derivatives of the stress warping functions $\psi'_{\alpha1}$ must also be continuous, which gives 2 more conditions per interface. Hence, for a $n$-layered plate, there are $2n+2n+3+3+=4n+6$ unknown constants and $4+2+2+2(n-1)+2(n-1)=4n+4$ equations. The same remark as above, concerning the solving procedure, applies here.
\par
The second subsystem is formed with the last two lines of system~\eqref{eq:FinalSystem1} and permits to compute $\psi_{22}$ and $\psi_{12}$ with the same procedure. It is interesting to remark that for these cases, the above process always gives $\varphi_{12}(z)=\varphi_{21}(z)=0$.
\subsection{Multilayered angle-ply othotropic plates}
The above procedure is applied. It gives non null $\varphi_{12}$ and $\varphi_{21}$ functions. The choice for $\gamma^0_{13}$, $\gamma^0_{13,1}$, etc. is not as obvious as for the cross-ply case. The functions that are built with this process seem to have good shapes, but it is difficult to verify them with finite element simulations because boundary conditions are difficult to choose. All the previous finite element computations were cylindrical bending of the plate in the $x$ or $y$ directions. For off-axis plates, it is always possible to enforce cylindrical bending but coupled effects makes the results difficult to interpret. Further studies are needed to know what is the better way to determine warping functions in this case. 

\section{Results}\label{sec:Results}
\subsection{The homogeneous --single layer-- case}\label{sec:homogeneous}
Here are presented the warping functions for the homogeneous case. We shall distinguish the isotropic and the orthotropic case for two reasons. The most evident reason is that, for an orthotropic material which is considered off-axis, four non-null warping functions are expected, instead of only two (different) for the on-axis configuration, and one (two identical) for the isotropic case. But there is another reason, also very interesting. For this material, it is possible to give to the shear transverse modulus $G_{13}$ and $G_{23}$ arbitrary (small) values. We are going to see that this has an influence on the shape of the warping functions. 
\subsubsection{The isotropic single layer case}\label{sec:isotropic_single}
\begin{figure}[ht]
	\centering
		\includegraphics[width=10cm]{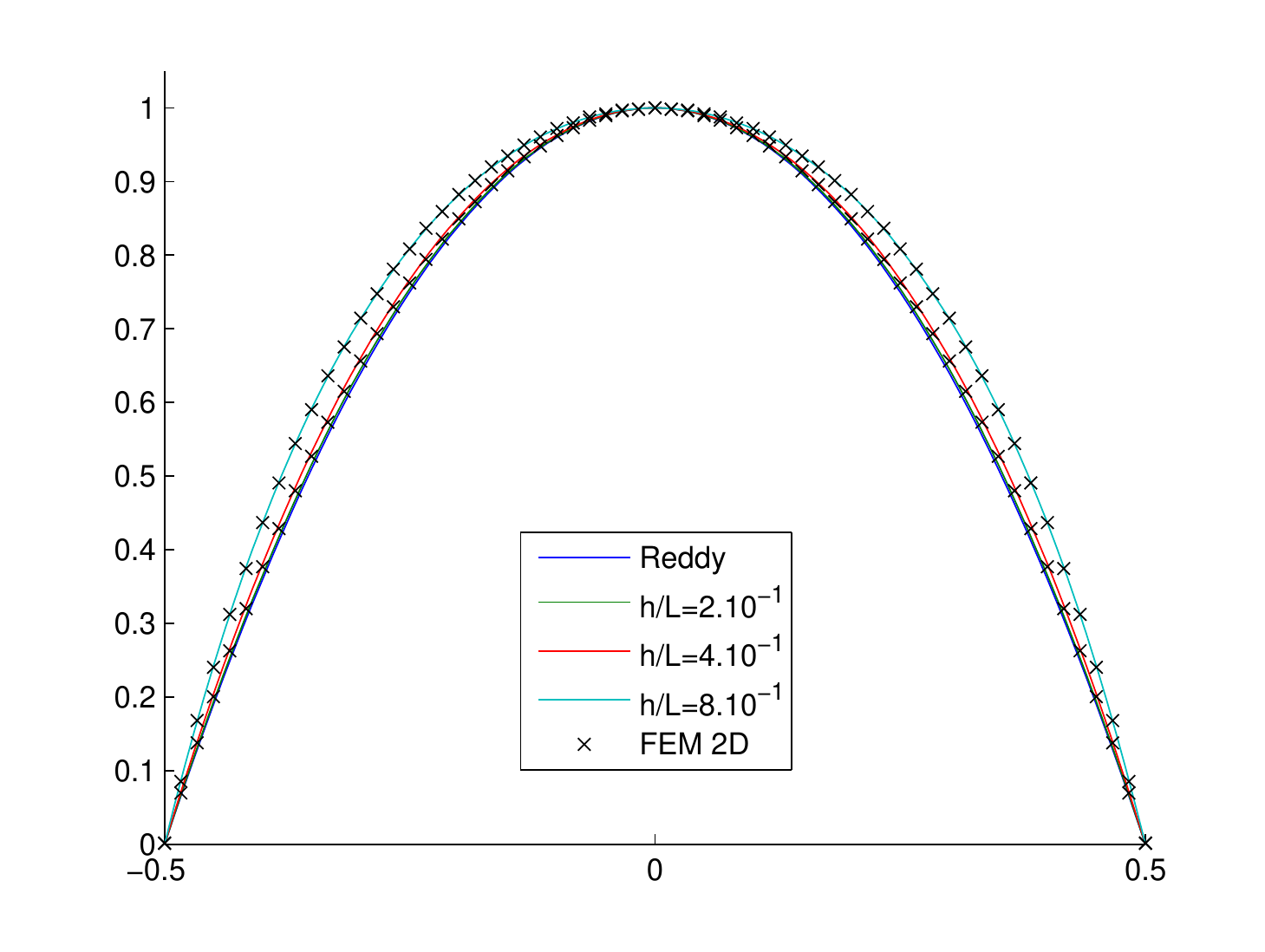}
	\caption{Warping functions for the case of a single layer isotropic laminate, with $\nu=0.3$ and $h/L$ varying from $2.10^{-1}$ to $8.10^{-1}$ compared to Reddy's formula $1-4(z/h)^2$. The two first curves are superimposed. For clarity, the 2D FEM simulation has been plotted only for the two higher values of $h/L$.}
	\label{fig:IsotropeElancement}
\end{figure}

An isotropic material with $\nu=0.3$ is considered here. As we shall see in figure~\ref{fig:IsotropeElancement}, the thickness to length ratio $h/L$ has an influence on the warping functions shapes. However Reddy's formula is a very good approximation until this ratio reaches the value of $4.10^{-1}$ which is uncommon for plates studies because three dimensional effects generally occurs with such structures. It is the reason why Reddy's model has been extensively used and is found to give good results for many practical applications. But we shall see below that the sensitivity of the $h/L$ ratio is higher when orthotropic material are considered. 
\begin{figure}[ht]
	\centering
		\includegraphics[width=10cm]{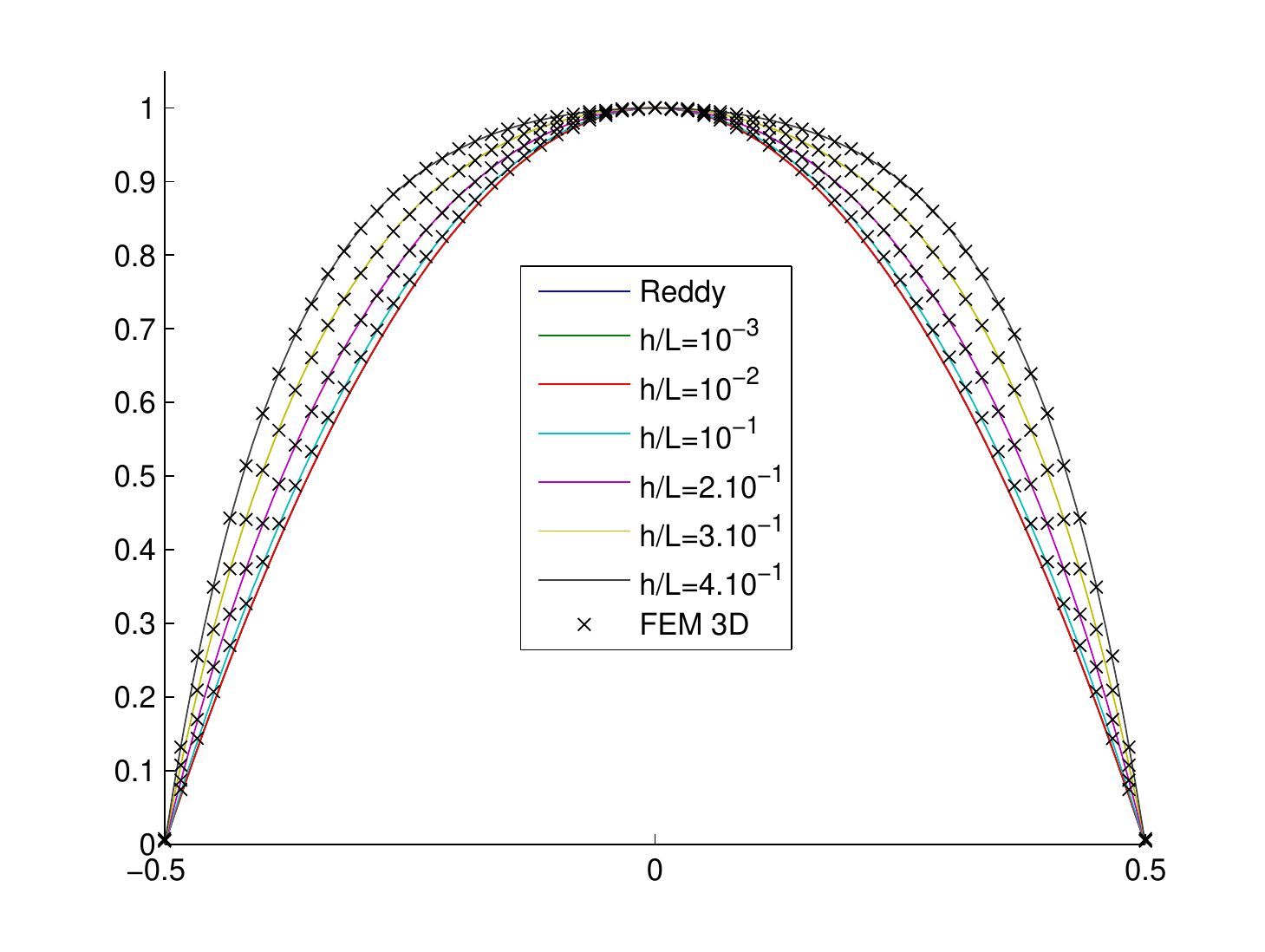}
	\caption{Warping functions for the case of a single layer orthotropic laminate with $G_{13}/E_1=40$ and $h/L$ varying from $10^{-3}$ to $4.10^{-1}$ compared to Reddy's formula $1-4(z/h)^2$. The three first curves are superimposed, and the fourth one only differs from Reddy's formula of about $0.02\%$. For clarity, the 3D FEM simulation has been plotted only for the four higher values of $h/L$.}
	\label{fig:Orthotrope40Elancement}
\end{figure}

\subsubsection{The orthotropic single layer case}\label{sec:OrthotropicSingle}
\textbf{Ratio $h/L$ varies:} The same study is performed for an orthotropic on-axis material for which the ratio $G_{13}/E_1$ has been set to $40$, which is a common value for an unidirectional composite ply. It shall be seen in figure~\ref{fig:Orthotrope40Elancement} that, as mentioned above, the $h/L$ ratio has more influence than for an isotropic material. 
\par
\begin{figure}[ht]
	\centering
		\includegraphics[width=10cm]{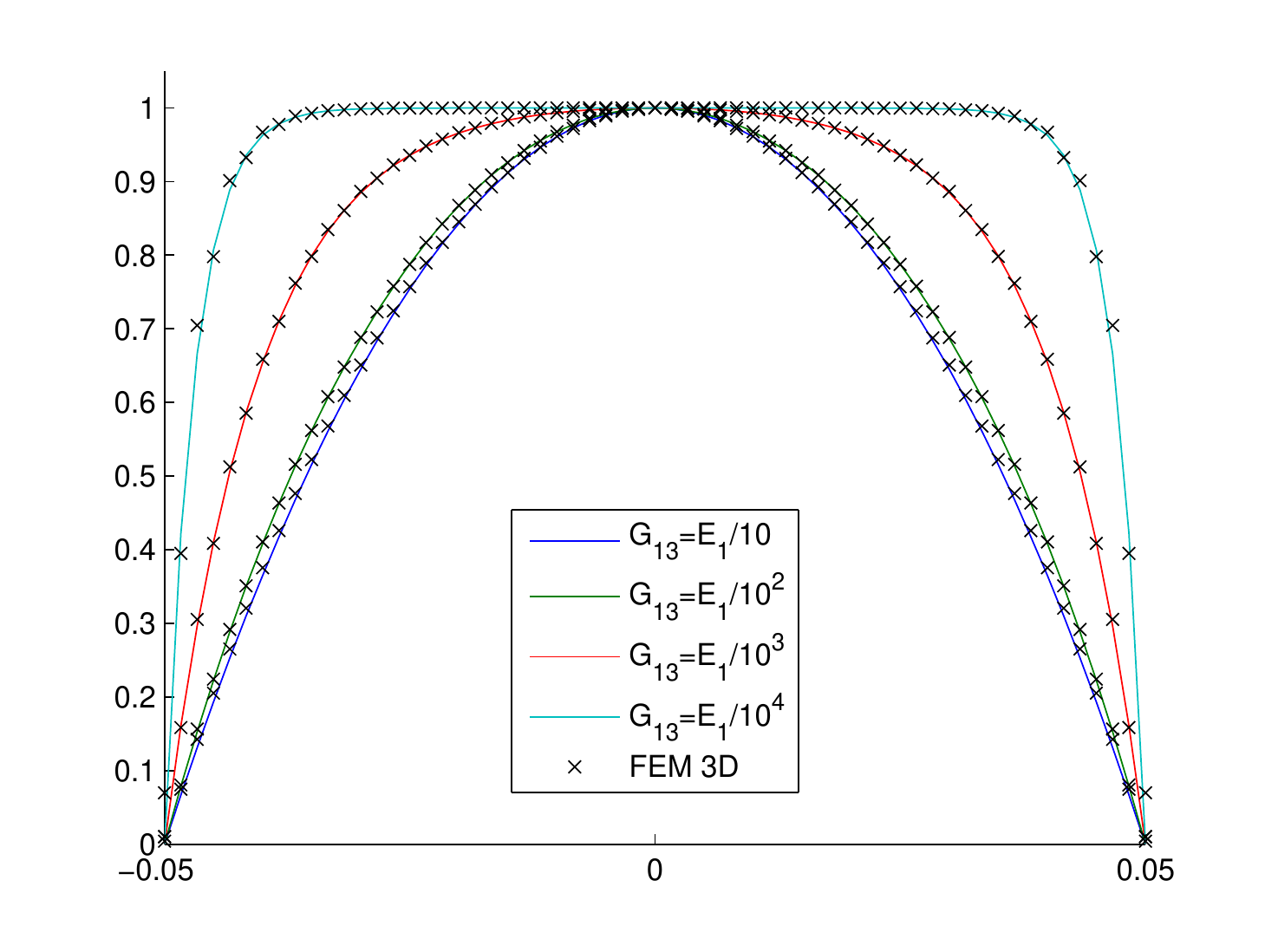}\;
	\caption{Warping functions for the on-axis orthotropic single layer case with $G_{13}$ varying from $E_1/10$ to $E_1/10^4$.}
	\label{fig:OrthotropeGVariable}
\end{figure}

\textbf{Ratio $G_{13}/E_1$ varies:} Now, an orthotropic on-axis material is studied for various values of ratio $G_{13}/E_1$ (see figure~\ref{fig:OrthotropeGVariable}). Other properties have been fixed to: $E_2=E_3=E_1$, $\nu_{23}=\nu_{13}=\nu_{12}=0.3$, $G_{23}=G_{13}$, $G_{12}=E_1/2(1+\nu_{12})$ and $h/L=0.1$. Figure~\ref{fig:OrthotropeGVariable} shows the warping functions, compared to 3D finite element computations.
\par
\begin{sidewaysfigure}
	\centering
		\includegraphics[width=5cm]{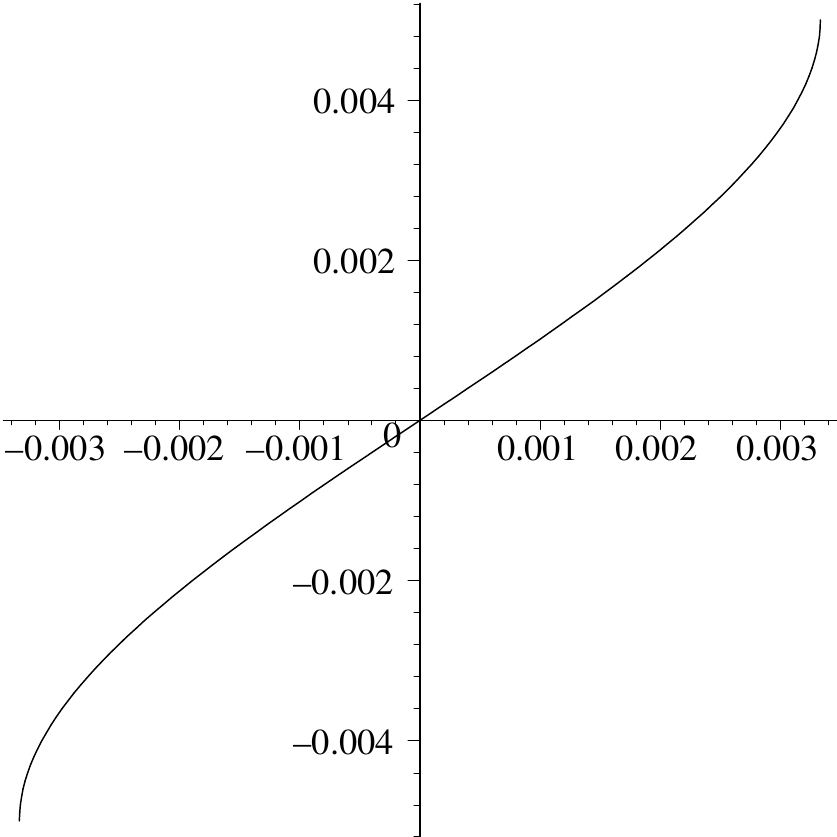}\;
		\includegraphics[width=5cm]{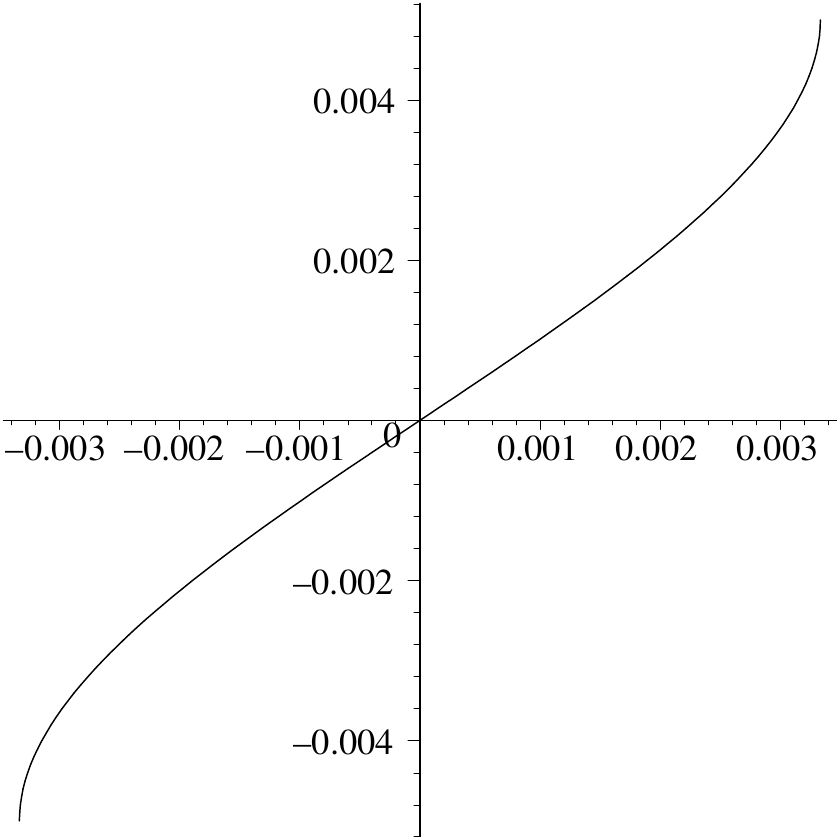}\;
		\includegraphics[width=5cm]{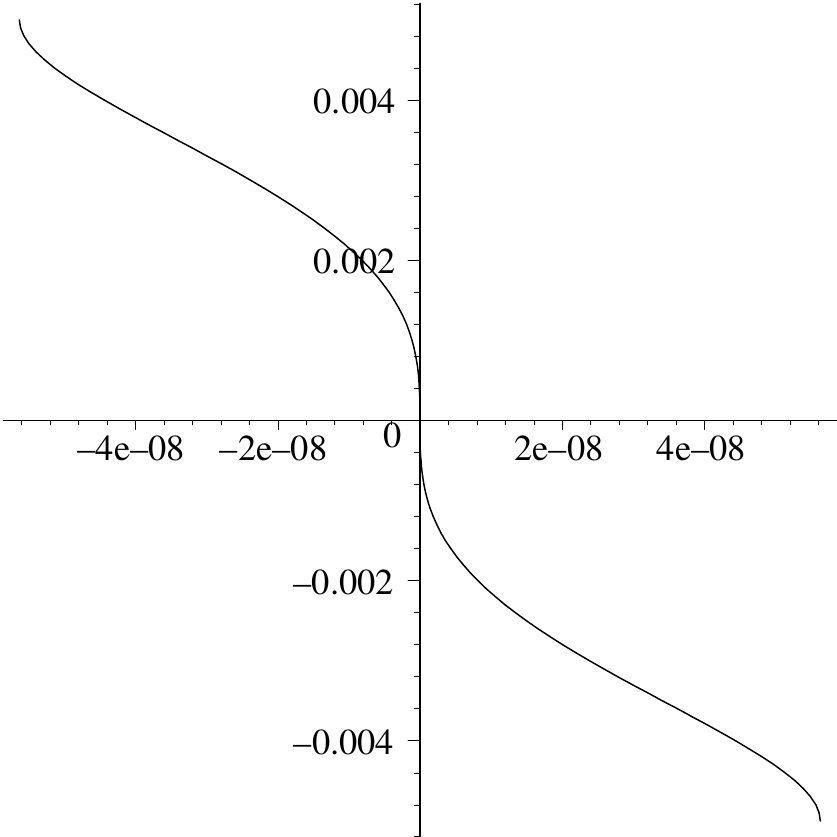}\;
		\includegraphics[width=5cm]{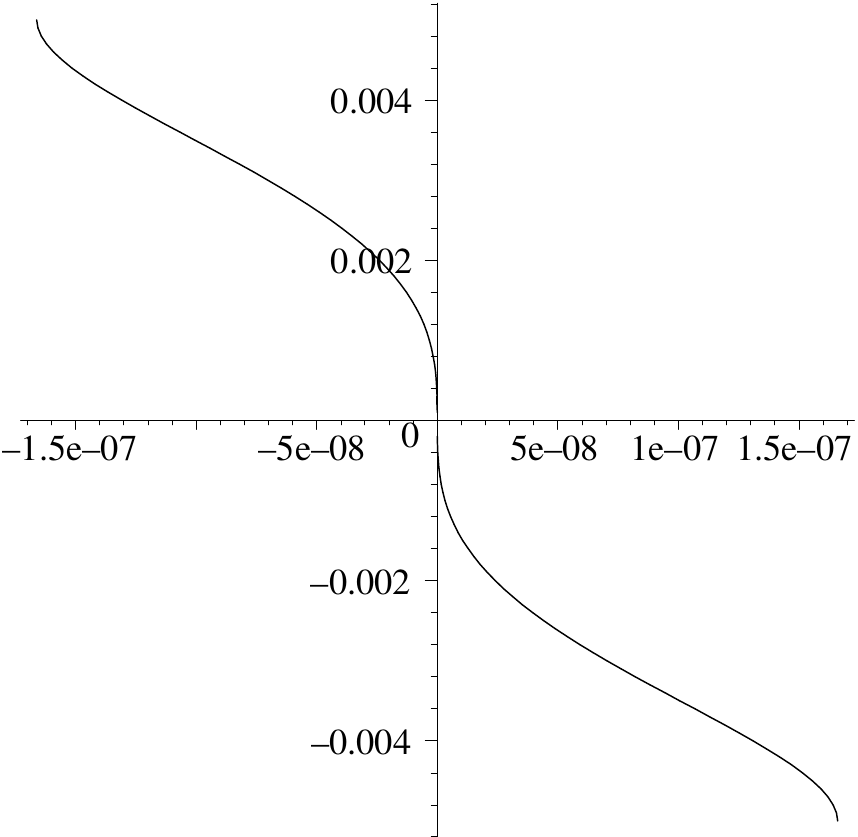}\\ \vskip 1 cm 
		\includegraphics[width=5cm]{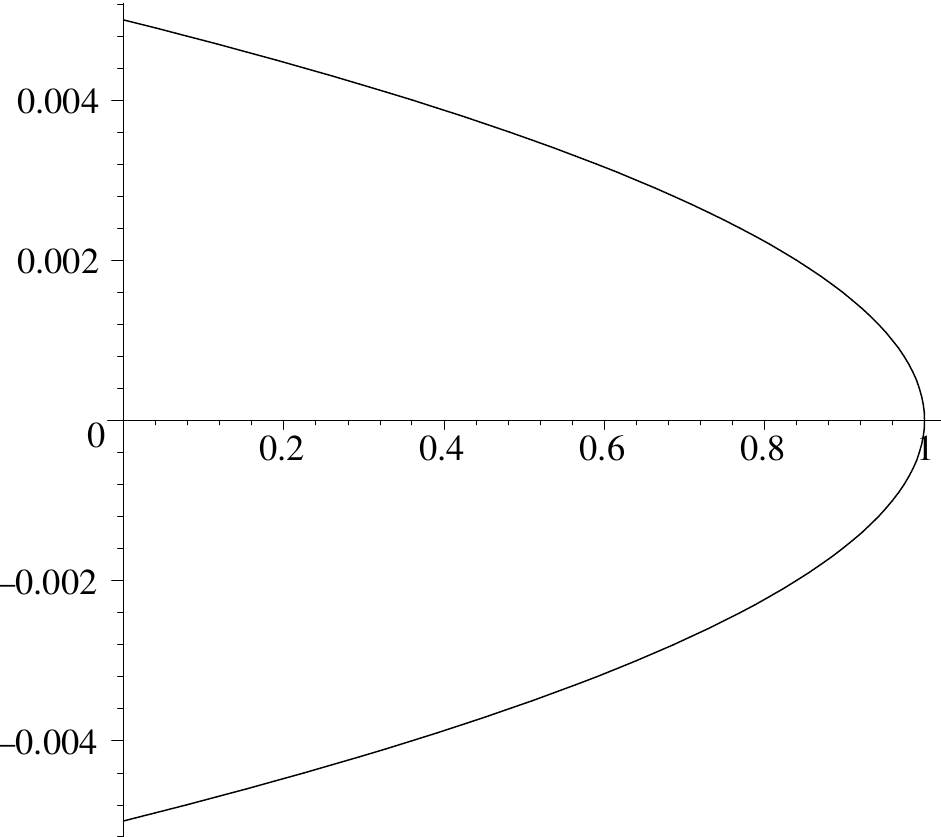}\;
		\includegraphics[width=5cm]{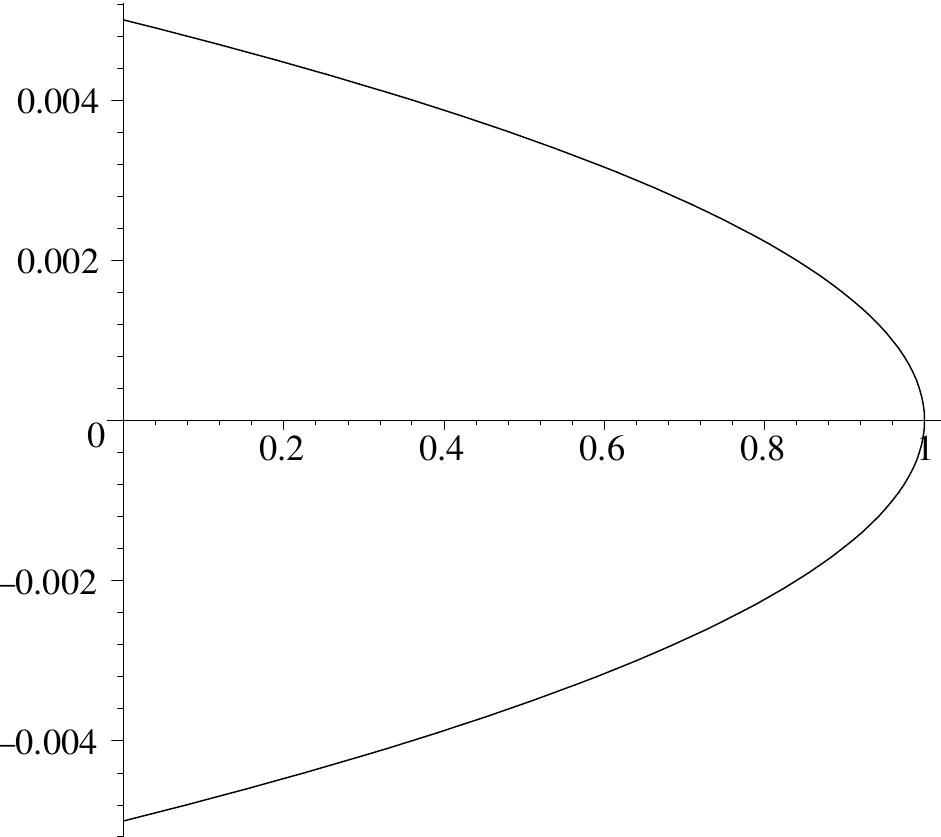}\;
		\includegraphics[width=5cm]{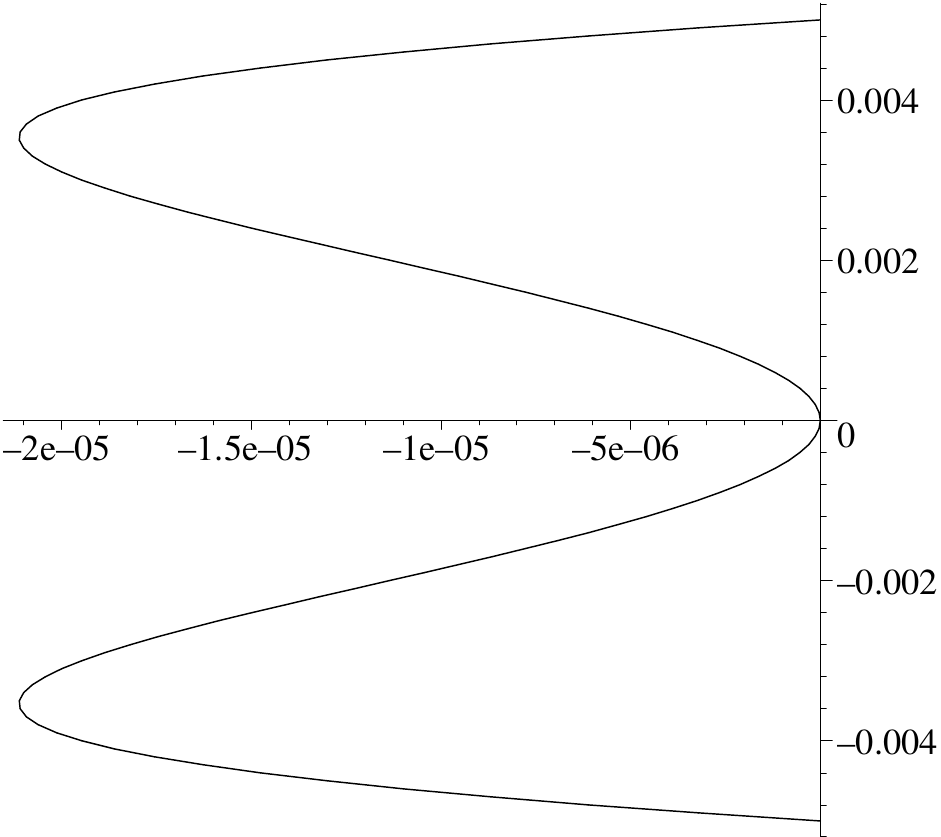}\;
		\includegraphics[width=5cm]{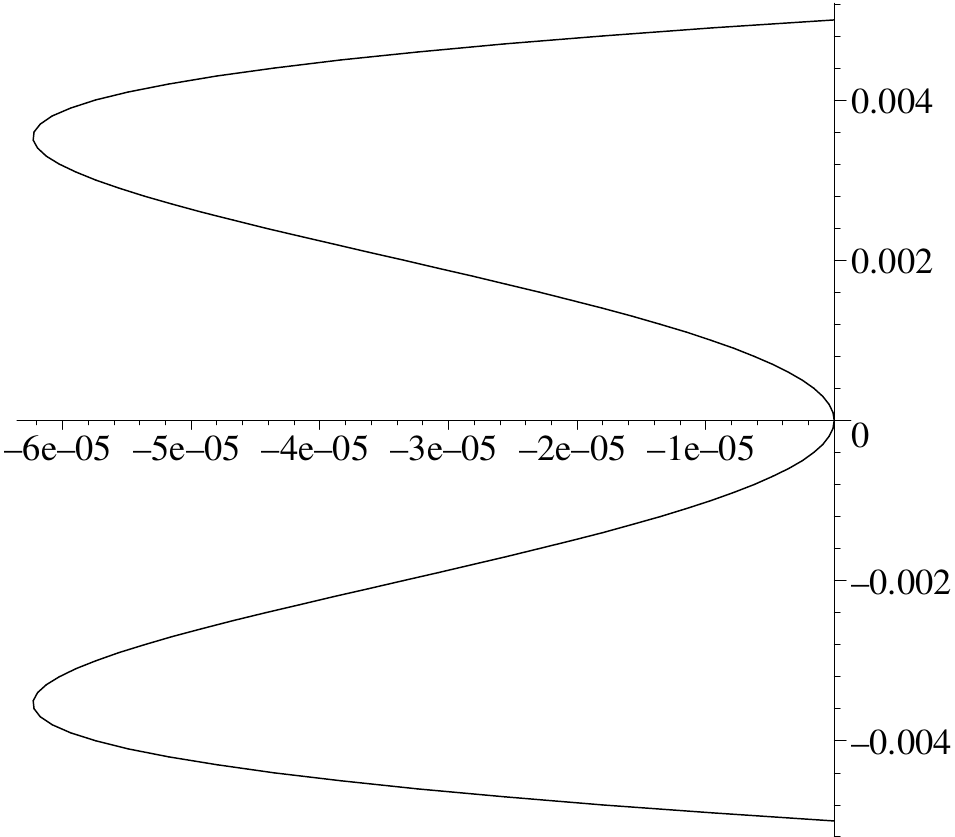}
	\caption{Warping functions and their derivatives for the case of an off-axis single layer orthotropic laminate with $\alpha = 30$}
	\label{fig:OrthotropicSingleLayer30}
\end{sidewaysfigure}

\textbf{Off-axis:} In this case, the choice of parameters $\gamma^0_{13}$, $\gamma^0_{23}$, $\gamma^0_{13,1}$, etc. is determinant. Taking values presented in the end of the section~\ref{sec:SolvingSystem} gives warping functions presented in figure~\ref{fig:OrthotropicSingleLayer30}. It is difficult to propose a finite element validation because a simple bending configuration have not yet been found.
%
%
%
%
%
\subsection{The multilayered case}\label{sec:multilayer}
\subsubsection{The isotropic multi layer case}\label{sec:IsotropicMulti}
\begin{sidewaysfigure}even layer
	\centering
		\includegraphics[width=5cm]{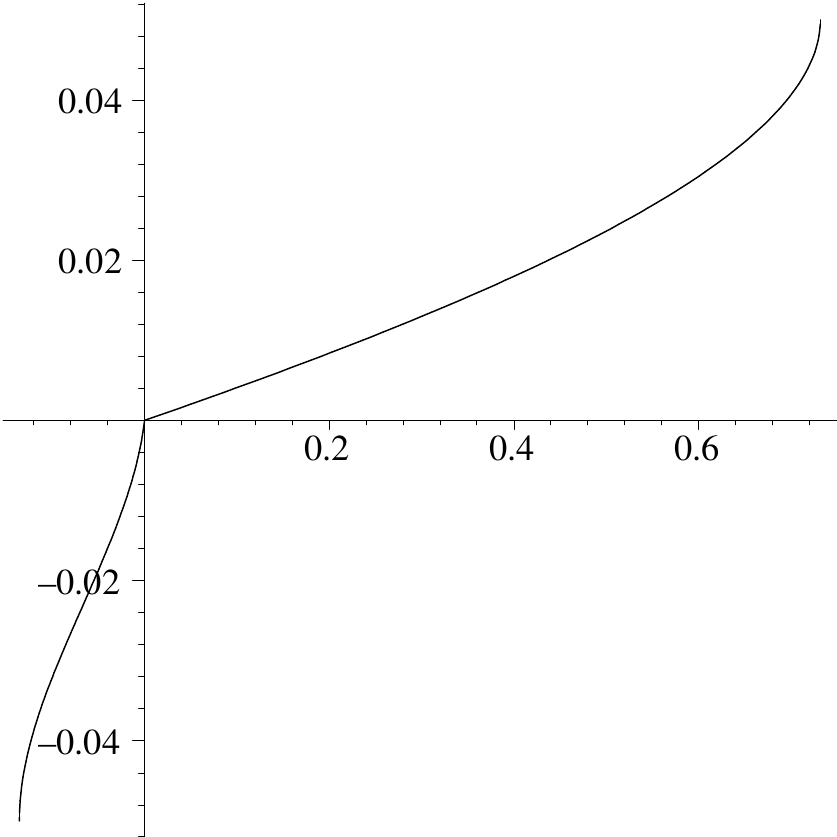}\;
		\includegraphics[width=5cm]{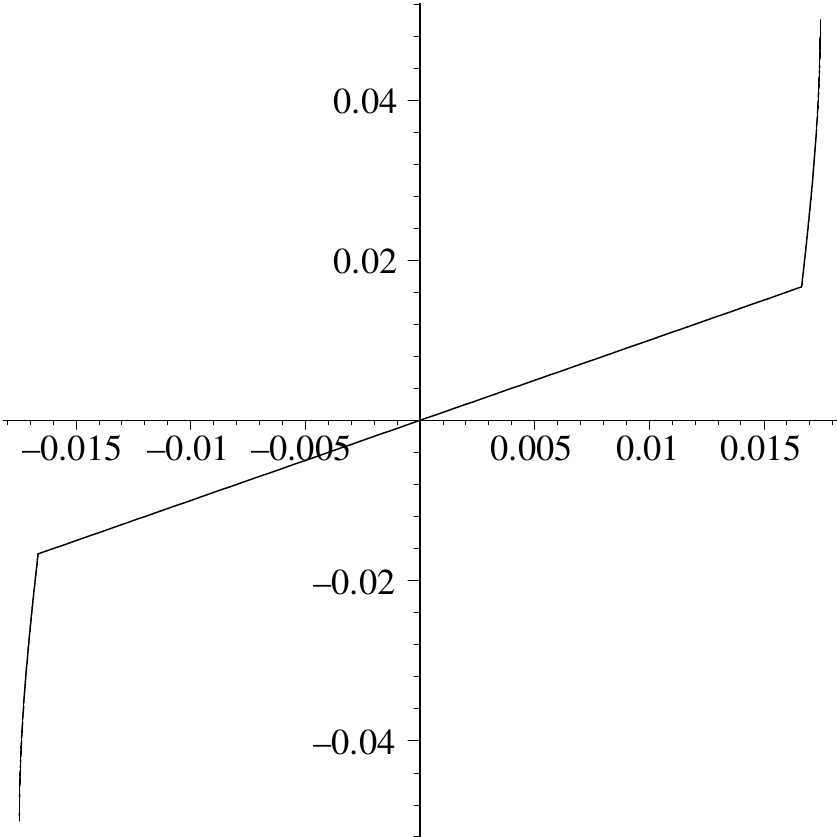}\;
		\includegraphics[width=5cm]{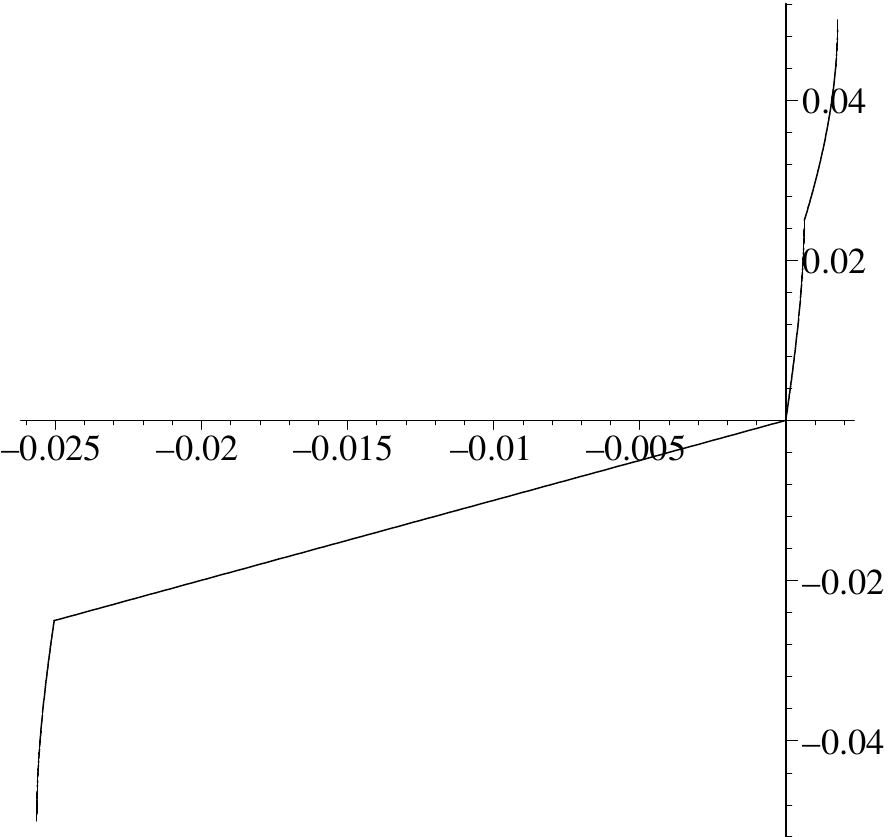}\; 
    \includegraphics[width=5cm]{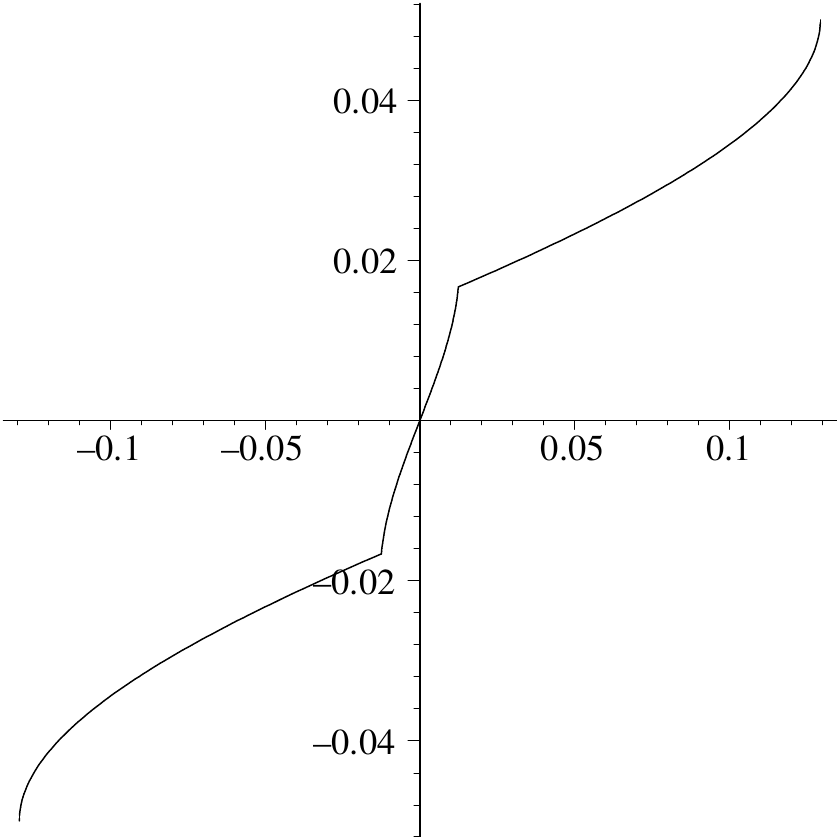}\; \vskip 1 cm
		\includegraphics[width=5cm]{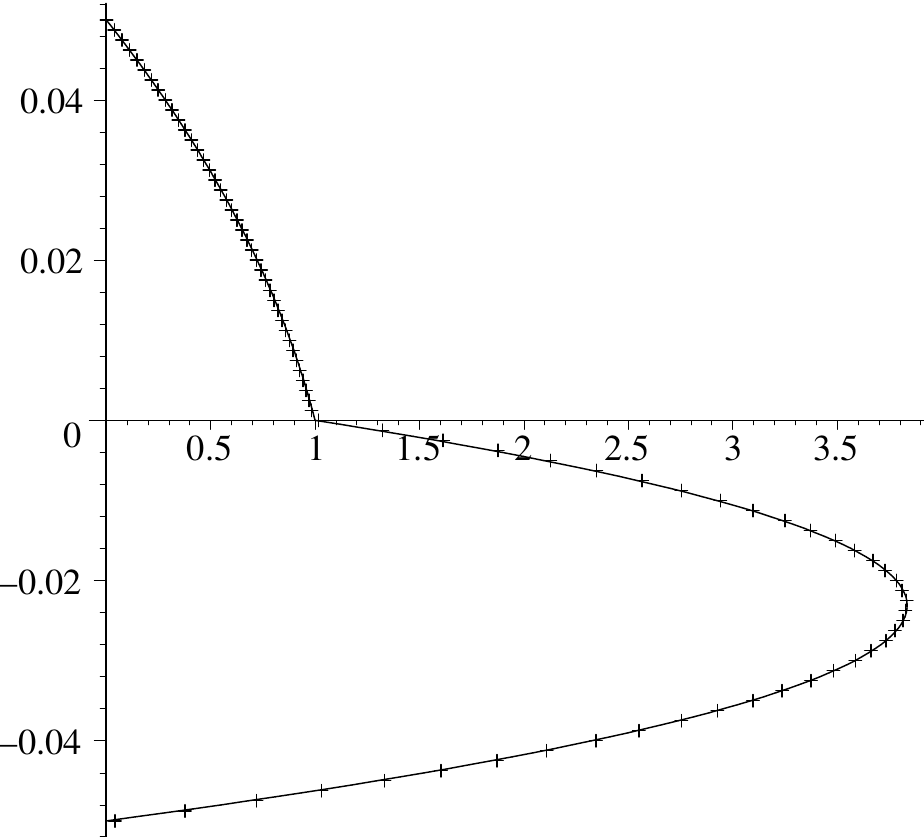}\;
		\includegraphics[width=5cm]{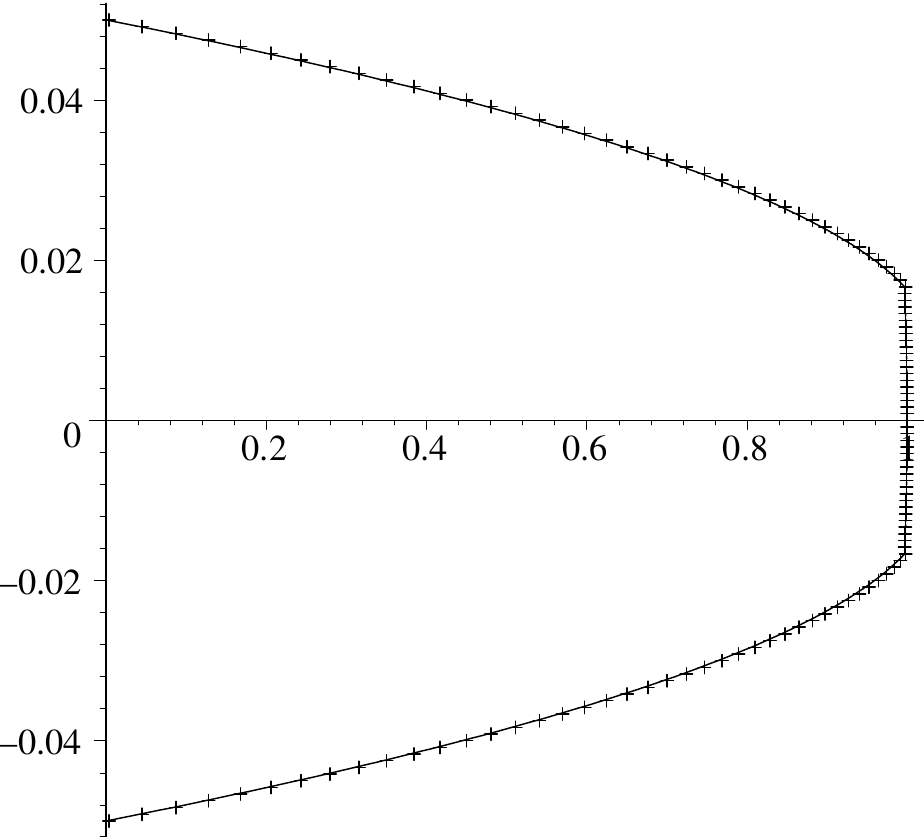}\;
		\includegraphics[width=5cm]{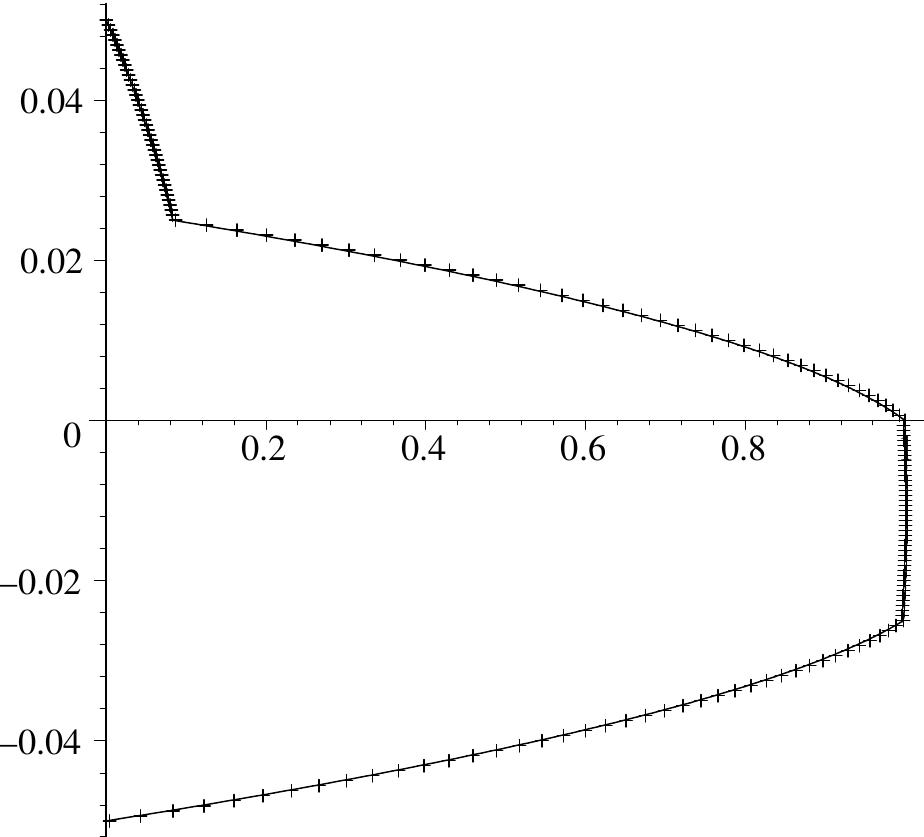}\;
		\includegraphics[width=5cm]{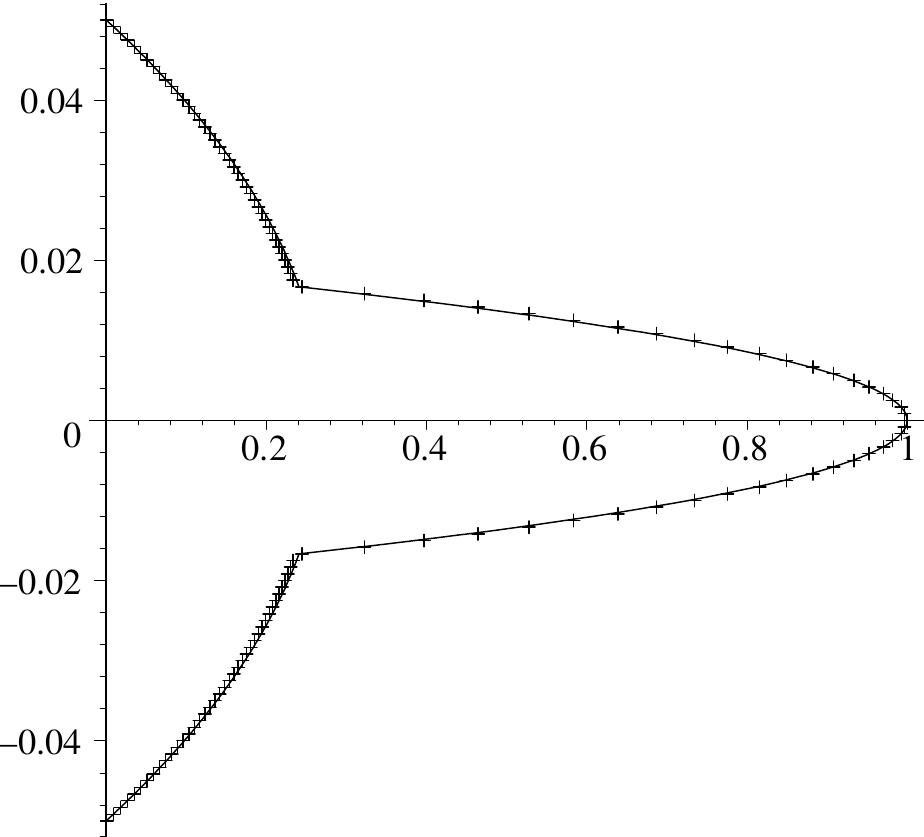}\;
	\caption{Warping functions and the corresponding transverse shear stress functions for the case of a n-layer isotropic laminate. From left to right: $n=2,3,4$ with $E_{\text{odd layer}} = 1/25 E_{\text{even layer}}$ and $n=3$ with $E_{\text{even layer}} =1/25 E_{\text{odd layer}}$. Cross correspond to 2D finite element computations.}
	\label{fig:IsotropicNLayer}
\end{sidewaysfigure}

In this section, a $n$-layer material with isotropic plies is studied. Young's modulus have been fixed according to the rule $E_{\text{odd layer}} =1/25 E_{\text{even layer}}$, except for the case $n=3$ for which the reciprocal $E_{\text{even layer}} =1/25 E_{\text{odd layer}}$ configuration is also given. Poisson's ratio is 0.3. The thickness to length ratio has been set to $h/L=0.1$. We shall see in figure~\ref{fig:IsotropicNLayer} the warping functions and their derivatives.
%
%
\subsubsection{The anisotropic multi layer case}\label{sec:AnisotropicMulti}
In this section, warping functions of some classical laminates are computed. Considered mechanical properties of the ply are: $E_1=400000$, $E_2=E_3=10000$, $\nu_{23}=\nu_{13}=\nu_{12}=0.25$, $G_{23}=6000$, $G_{13}=G_{12}=5000$ .   
\par 
\textbf{The $[0/90]$ laminate:} Unfortunately, at this time, the above procedure does not give correct warping functions. The problem could be due to coupling effects (this laminate has a membrane-bending coupling) but it is also the case for the isotropic two-layered laminate of section~\ref{sec:IsotropicMulti} for which warping functions were found and agreed very well. The problem could also be due to the discontinuity of the ratio between transverse shear and longitudinal modulus at the neutral plane. Further investigations need to be done for this configuration.
\par
\begin{sidewaysfigure}
	\centering
		\includegraphics[width=5cm]{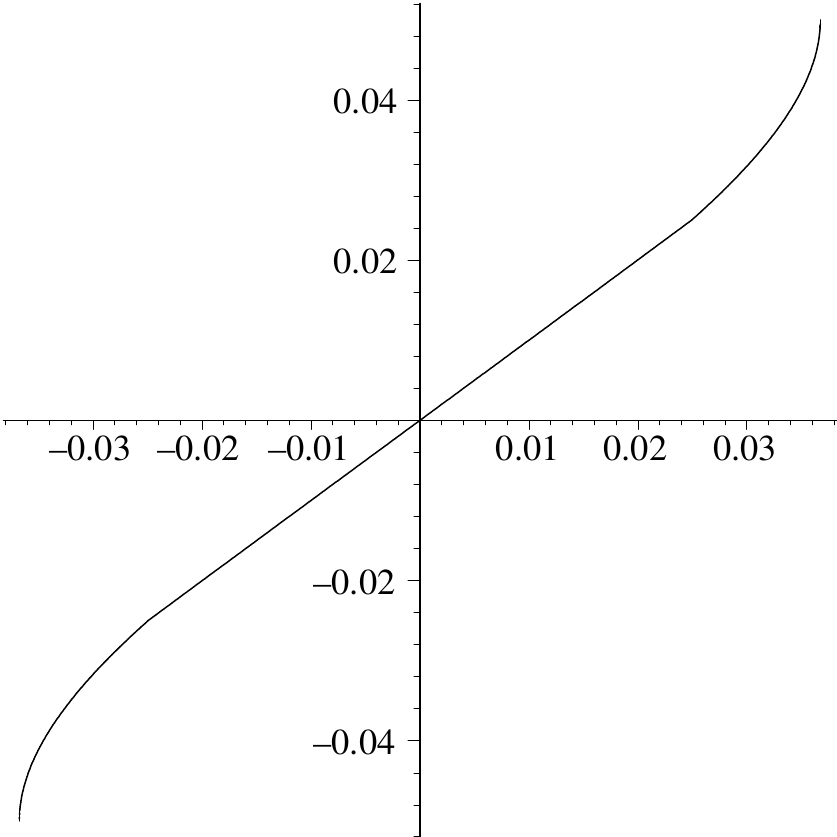}\;
		\includegraphics[width=5cm]{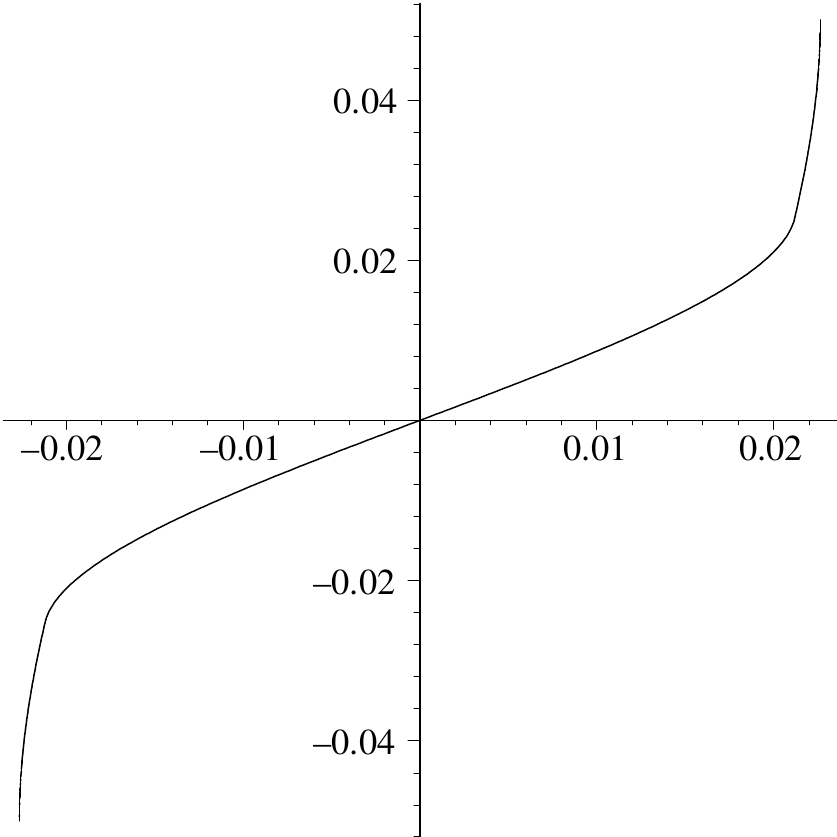}\;
		\includegraphics[width=5cm]{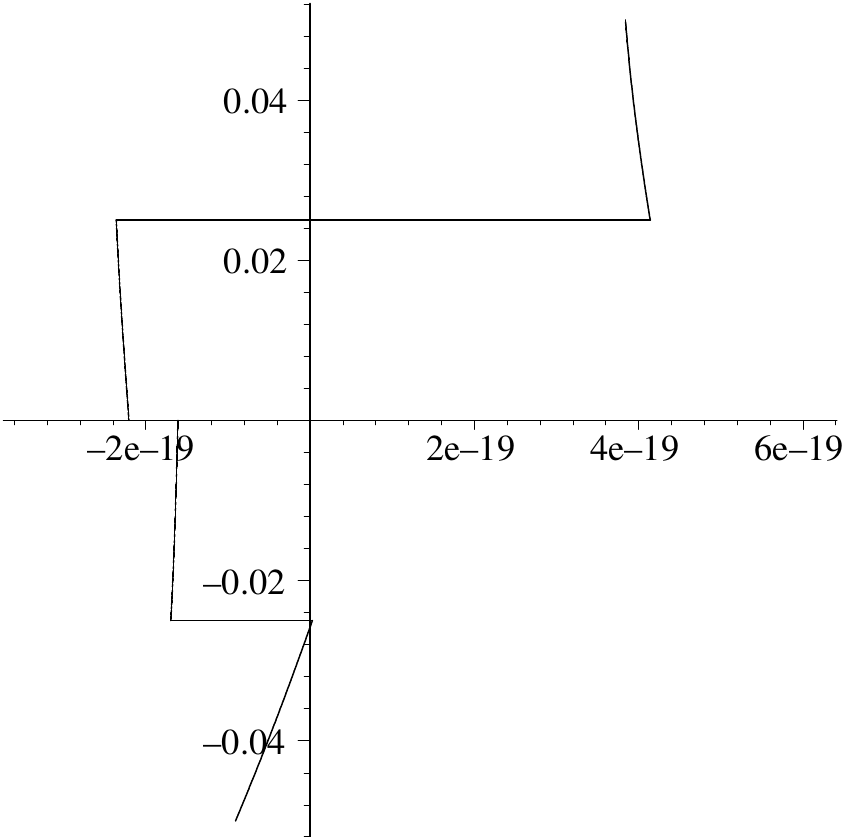}\;
		\includegraphics[width=5cm]{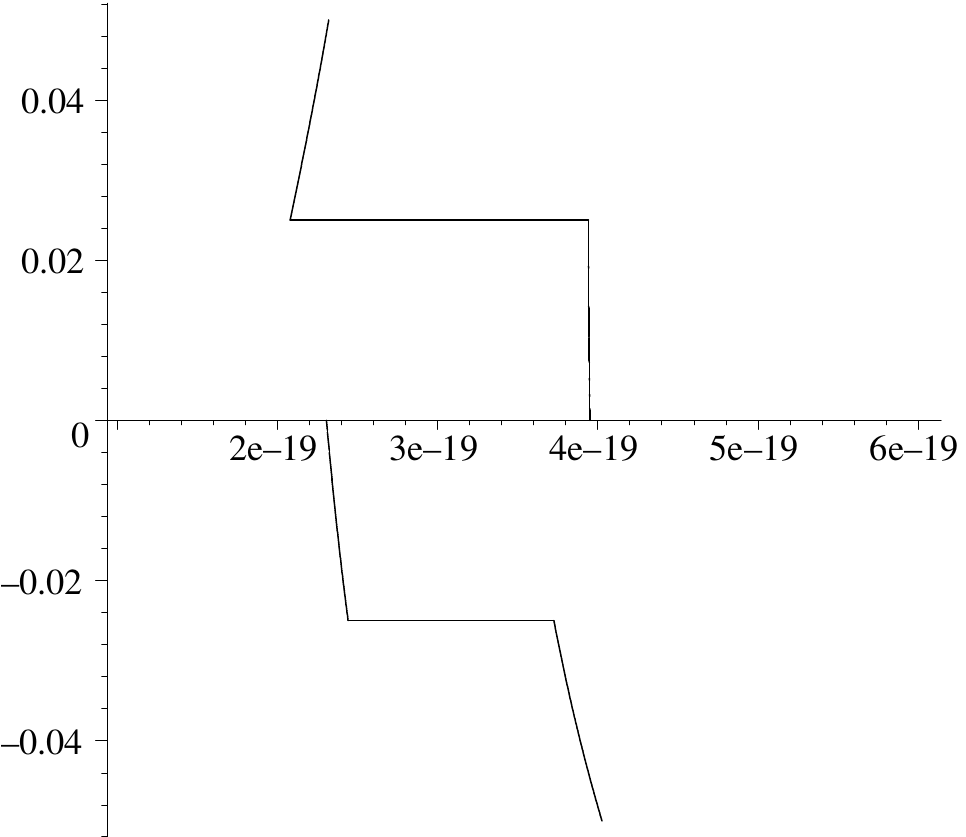}\\ \vskip 1 cm 
		\includegraphics[width=5cm]{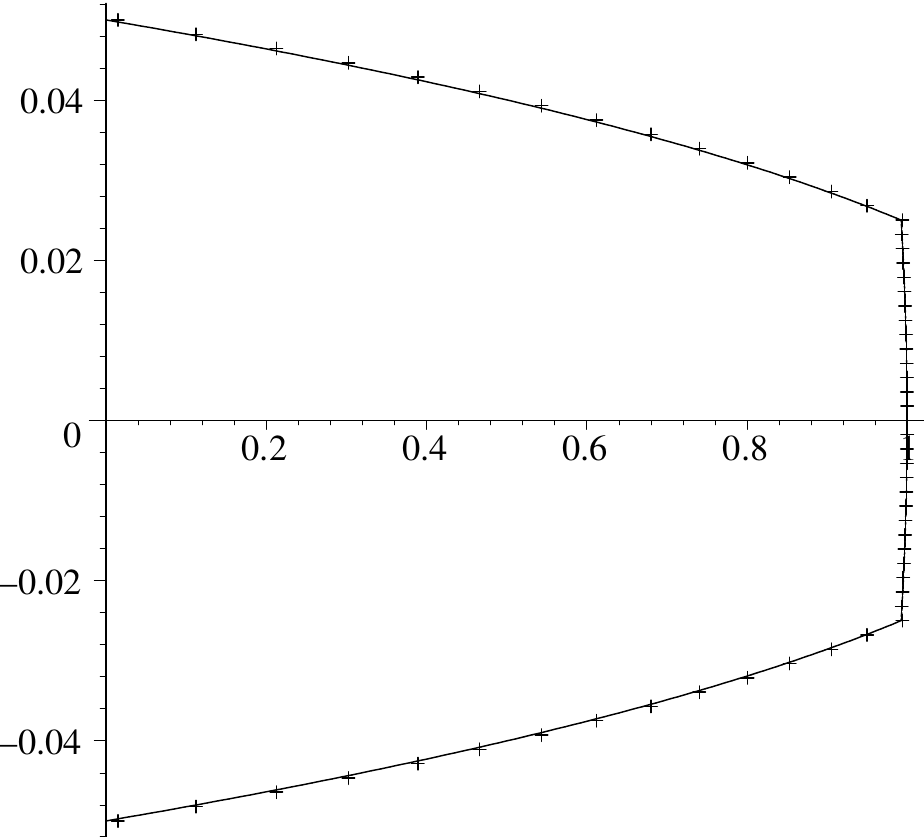}\;
		\includegraphics[width=5cm]{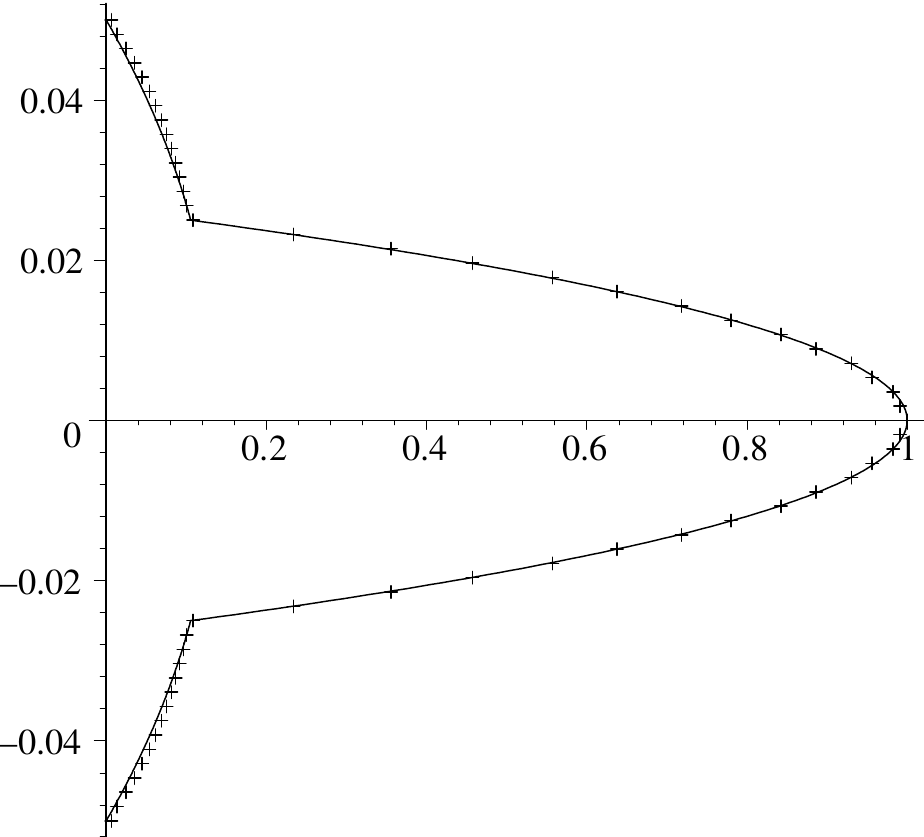}\;
		\includegraphics[width=5cm]{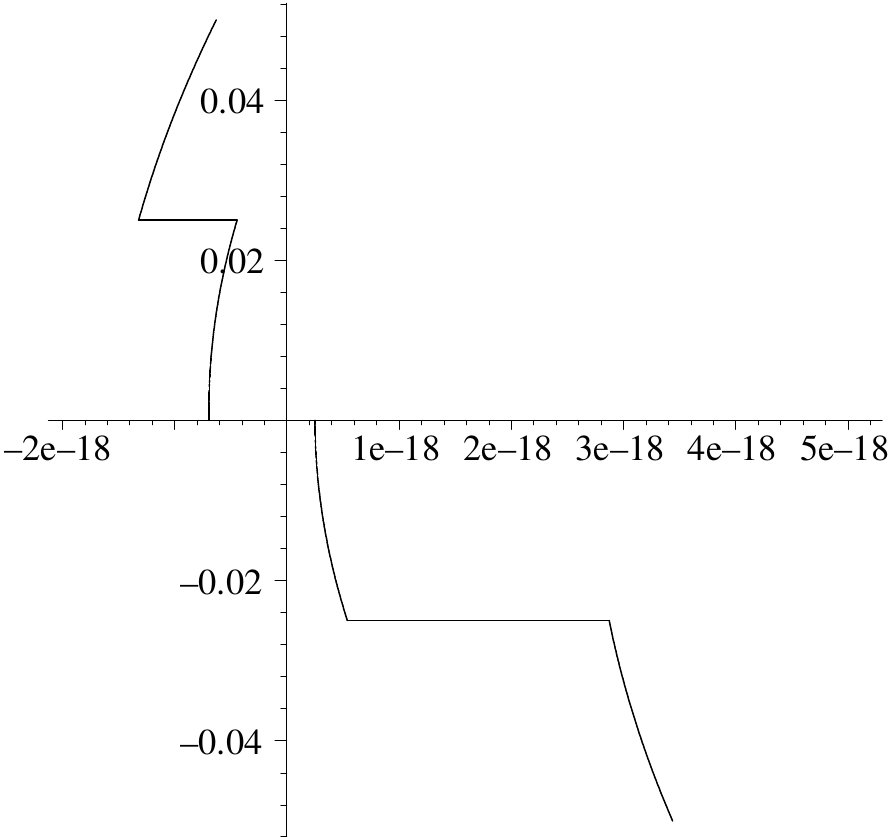}\;
		\includegraphics[width=5cm]{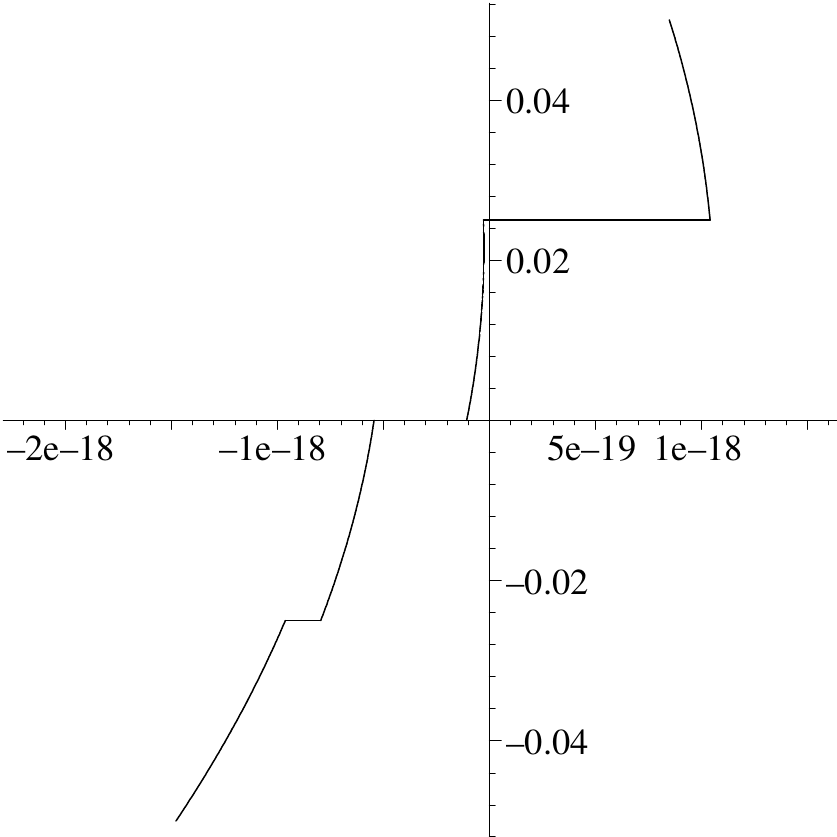}
	\caption{Warping functions and the corresponding transverse shear stress functions for the case of a $[0/90]_s$ laminate. Cross correspond to 3D finite element computations. Results for the $\varphi_{12}$, $\varphi_{21}$, $\psi'_{12}$ and $\psi'_{21}$ are intentionally left to show the behavior of the algorithm.}
	\label{fig:Orthotropic_0_90_90_0}
\end{sidewaysfigure}

 
\textbf{The $[0/90]_s$ laminate:} In this case, the cylindrical bending gives a decoupled strain state, and finite element results are in perfect agreement with warping functions computed from the present procedure, which can be seen in figure~\ref{fig:Orthotropic_0_90_90_0}.
\par
\begin{sidewaysfigure}
	\centering
		\includegraphics[width=5cm]{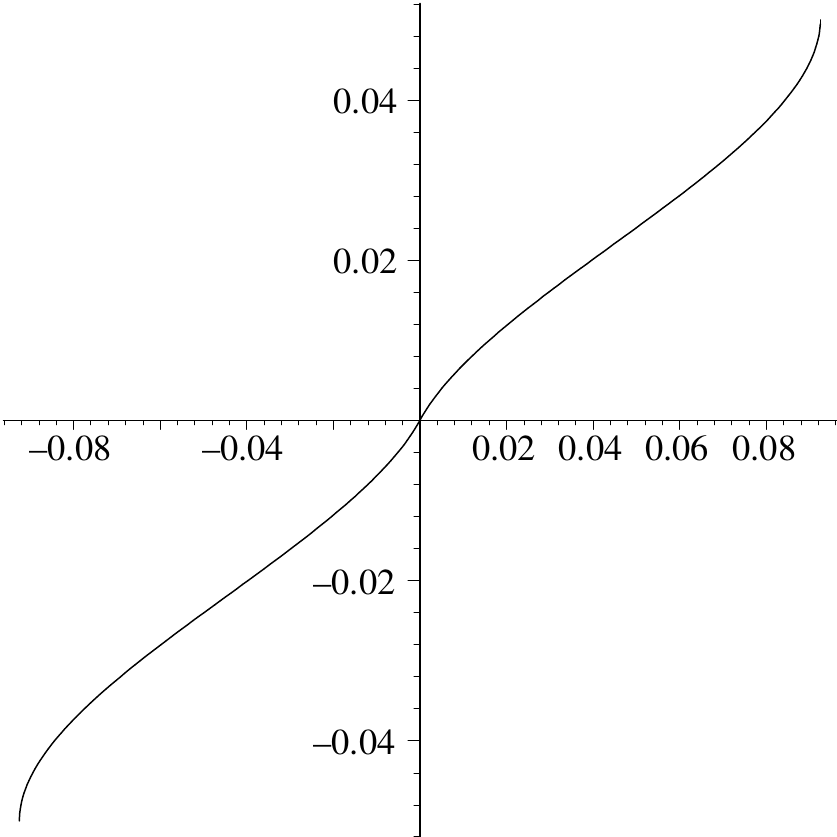}\;
		\includegraphics[width=5cm]{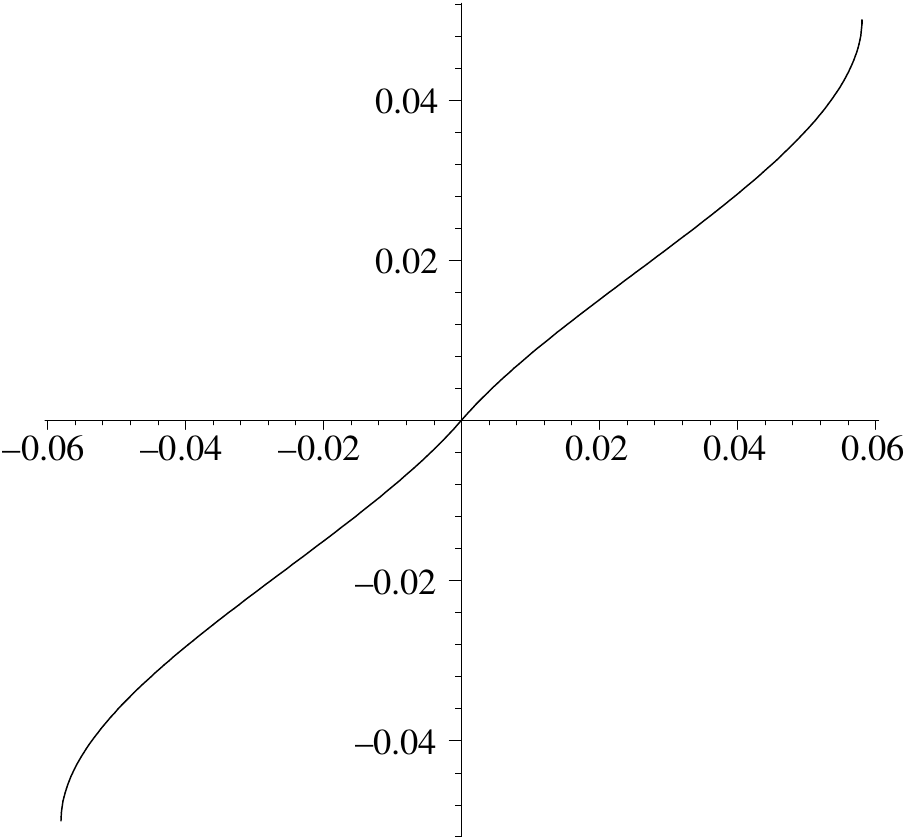}\;
		\includegraphics[width=5cm]{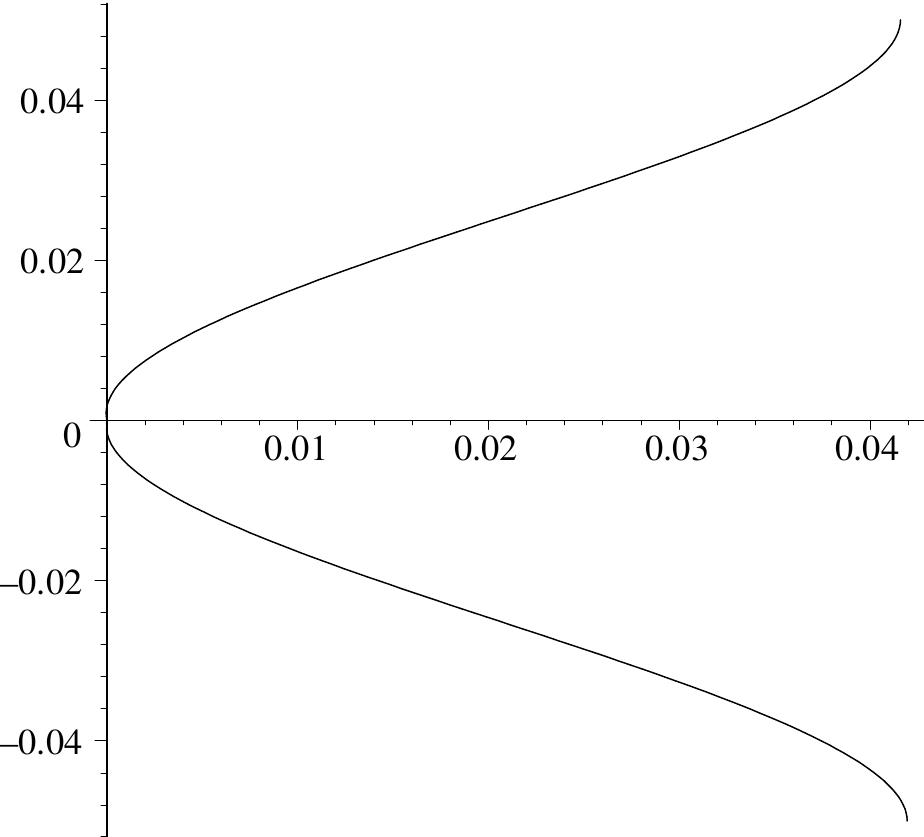}\;
		\includegraphics[width=5cm]{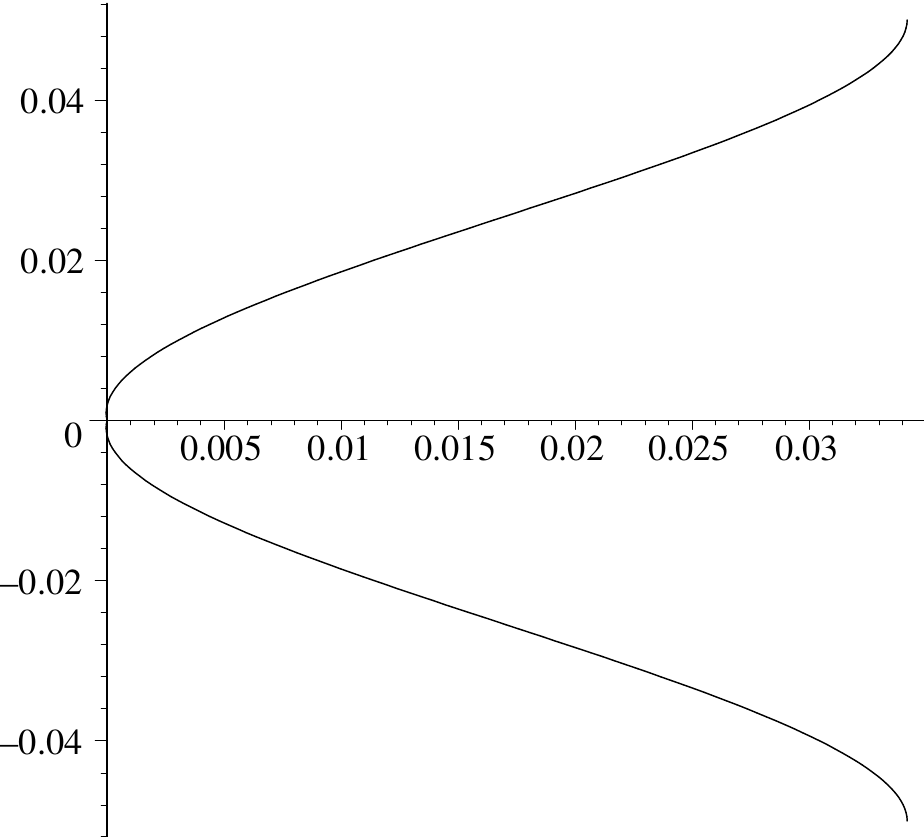}\\ \vskip 1 cm 
		\includegraphics[width=5cm]{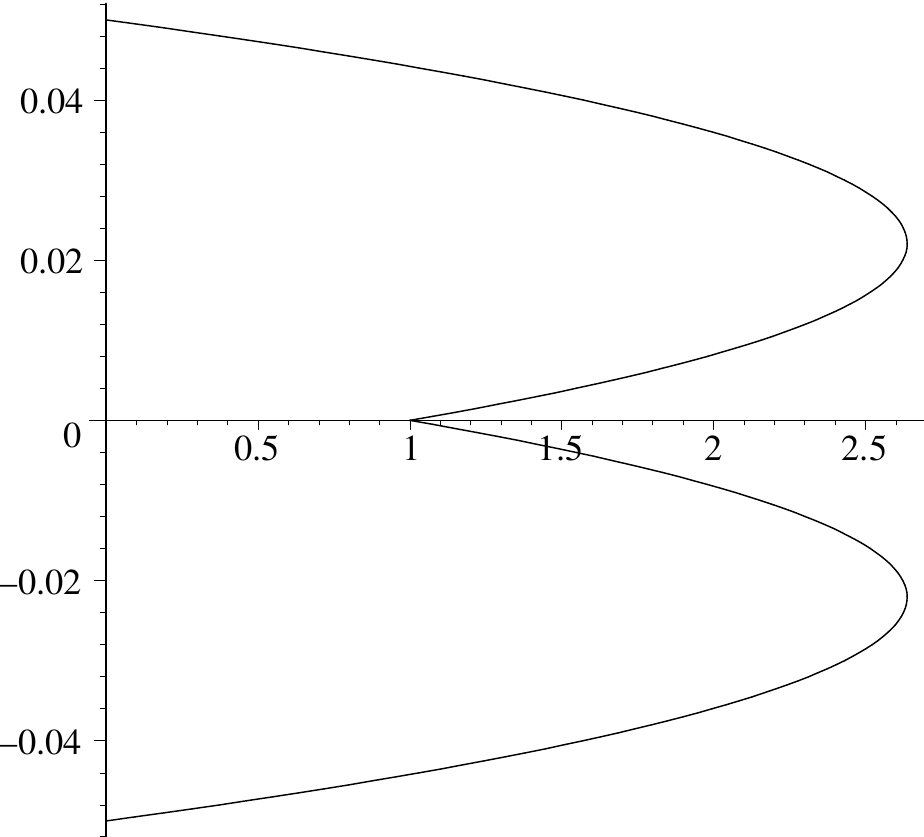}\;
		\includegraphics[width=5cm]{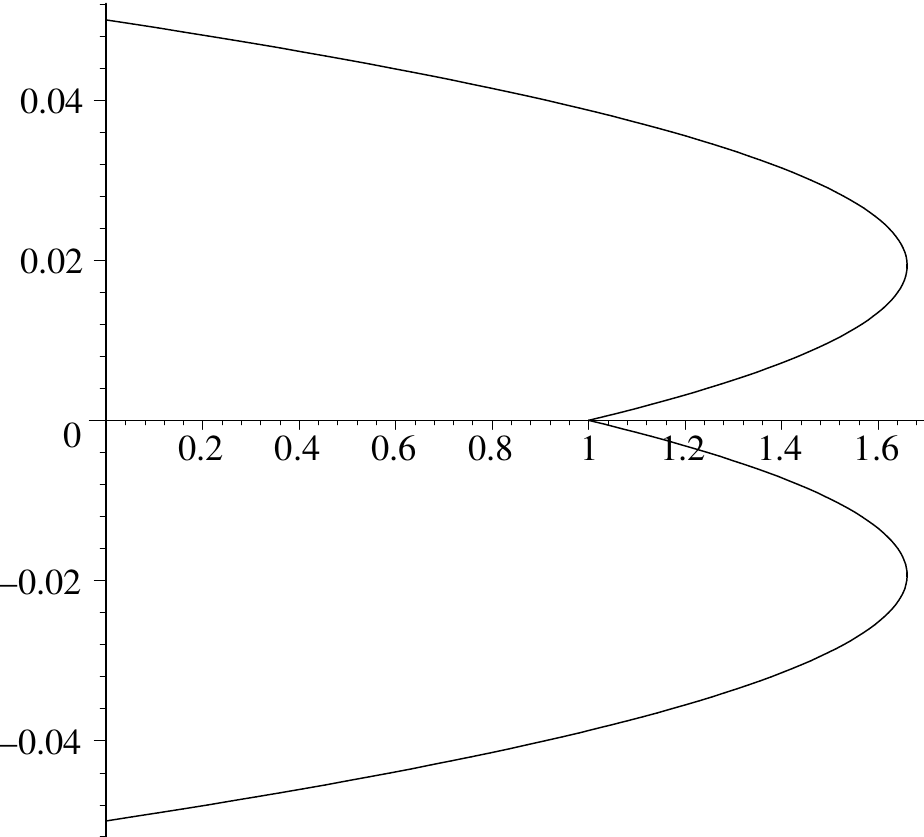}\;
		\includegraphics[width=5cm]{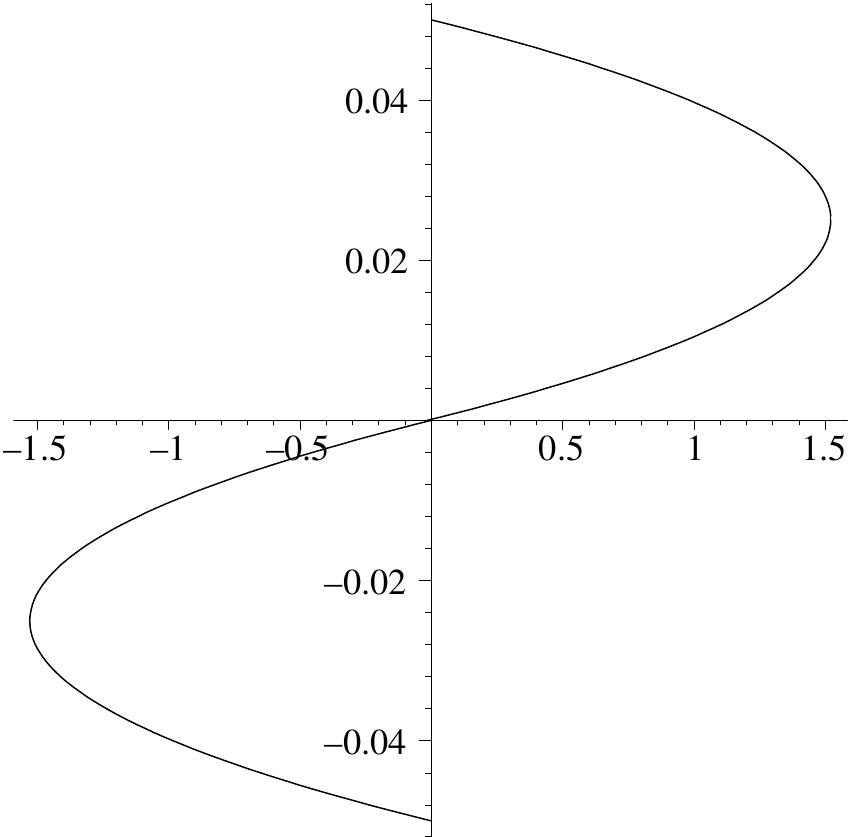}\;
		\includegraphics[width=5cm]{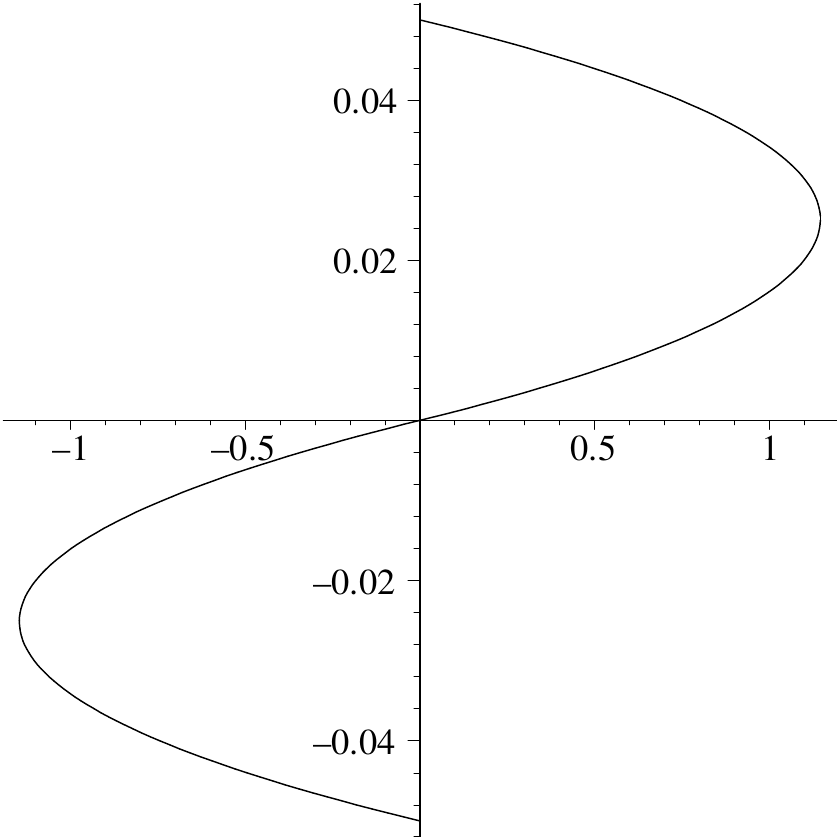}
	\caption{Warping functions and their derivatives for the case of a $[30/-30]$ laminate.}
	\label{fig:Orthotropic_p30_m30}
\end{sidewaysfigure}

\textbf{The $[30/-30]$ laminate:} It this case, the procedure gives warping functions which are plausible but no finite element proof can been given at this time for the same reasons than those of the $[30]$ one layer laminate of section~\ref{sec:OrthotropicSingle}. They are presented in figure~\ref{fig:Orthotropic_p30_m30}. 
\par
\begin{sidewaysfigure}
	\centering
		\includegraphics[width=5cm]{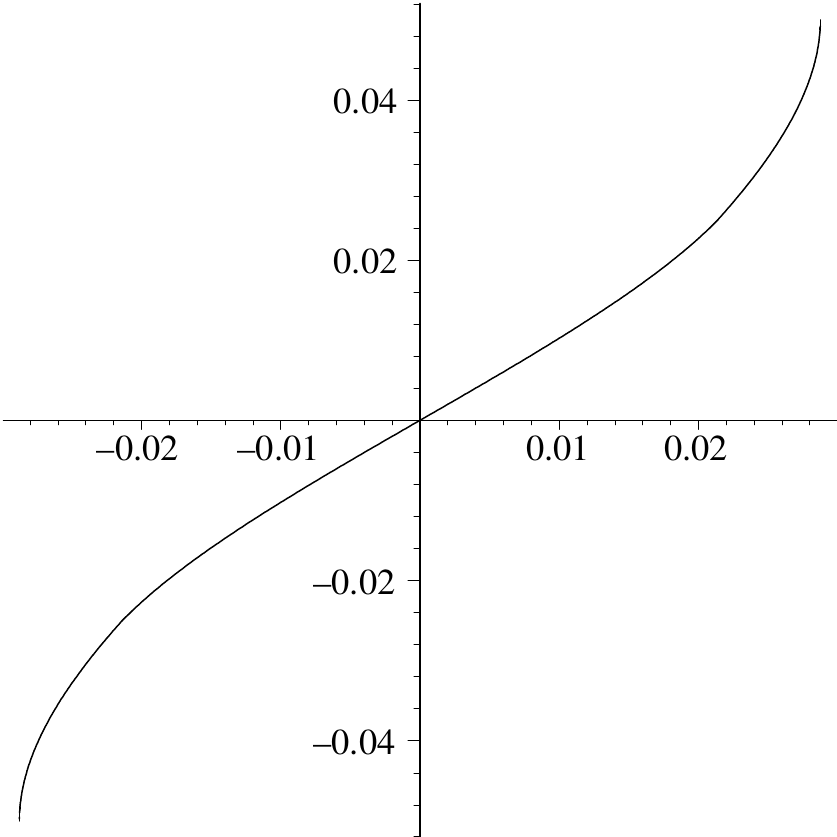}\;
		\includegraphics[width=5cm]{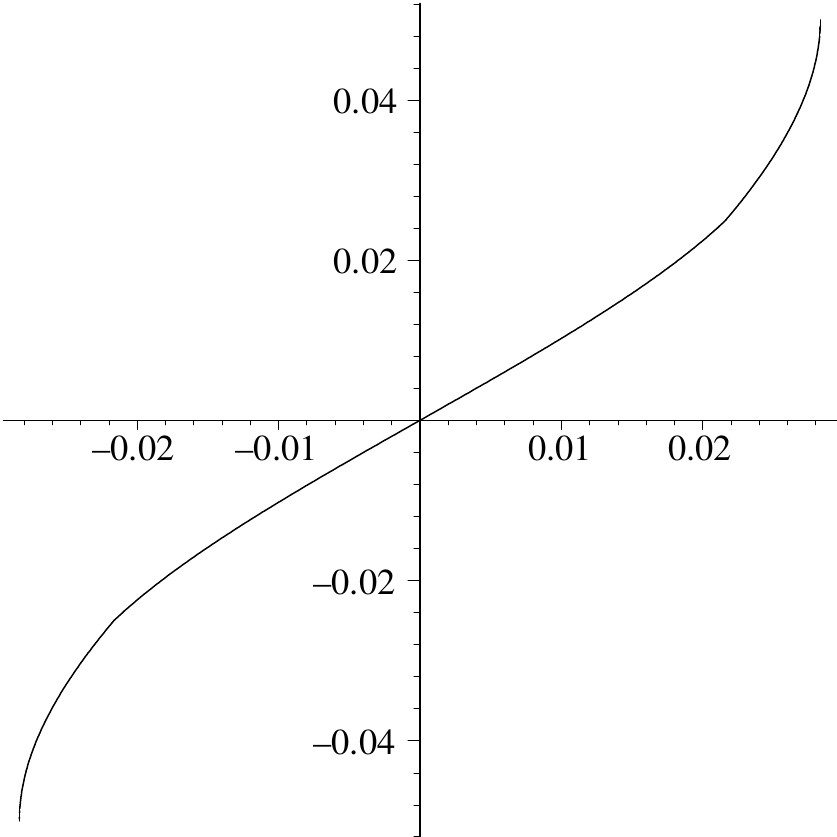}\;
		\includegraphics[width=5cm]{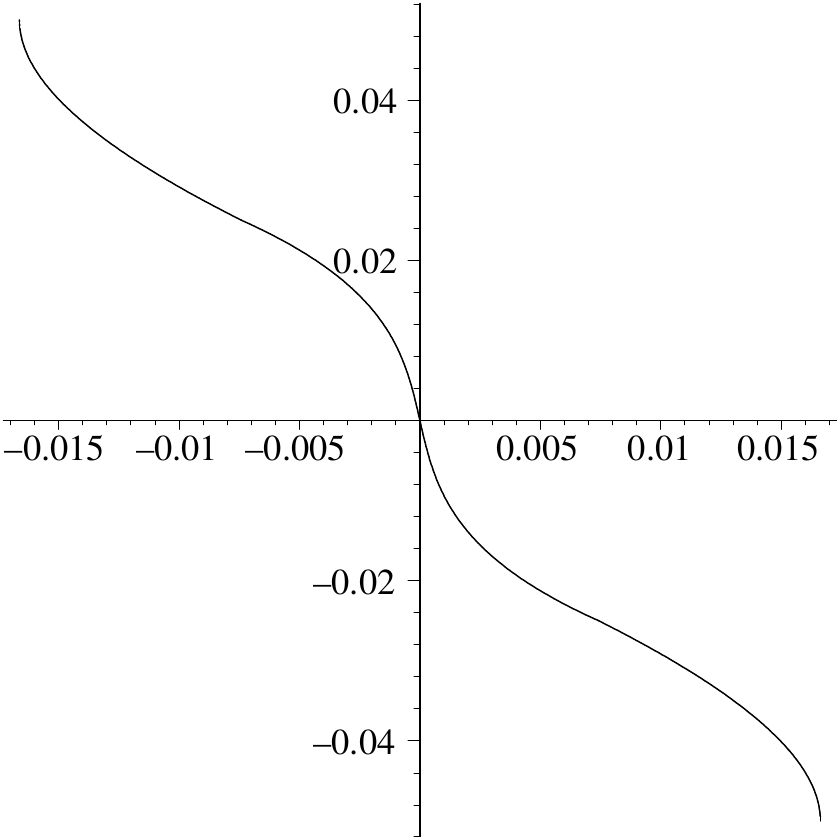}\;
		\includegraphics[width=5cm]{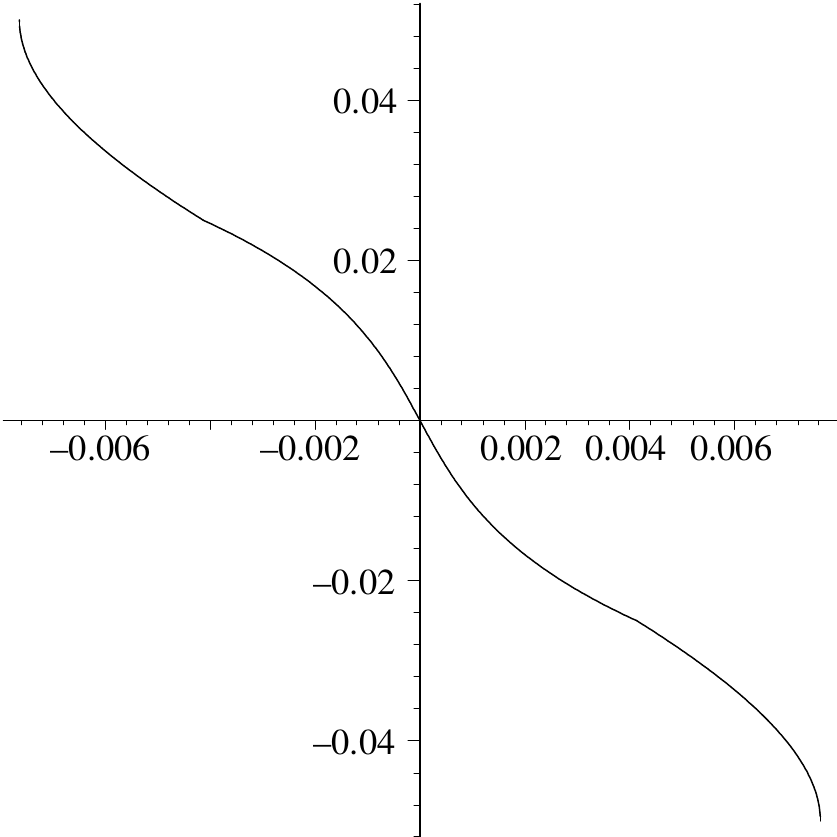}\\ \vskip 1 cm 
		\includegraphics[width=5cm]{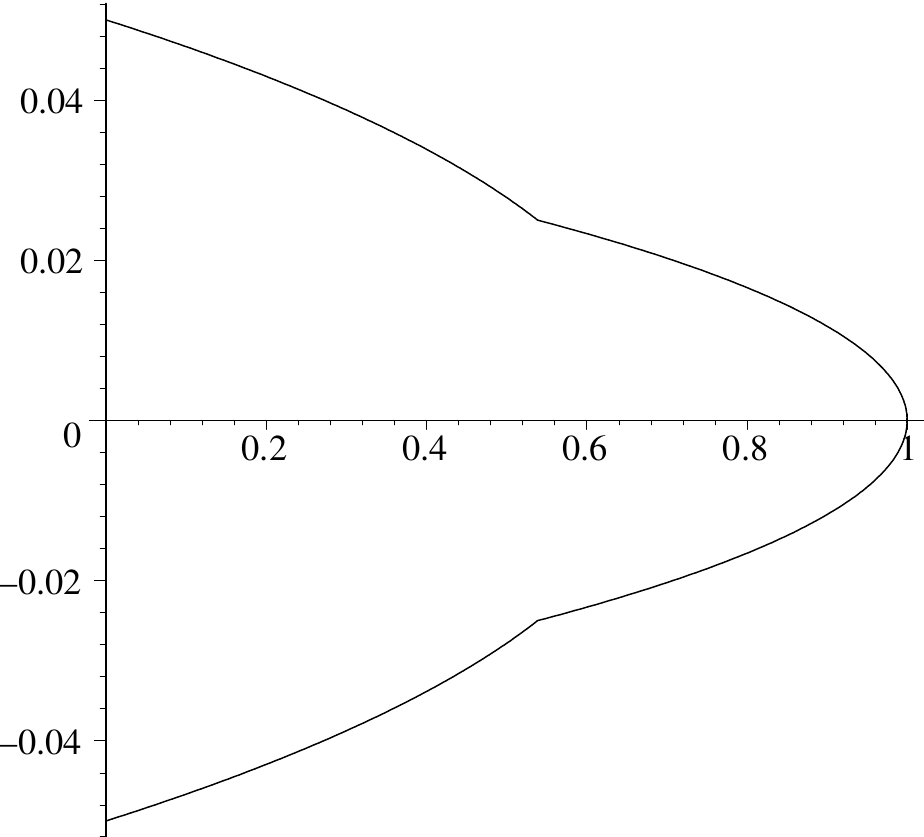}\;
		\includegraphics[width=5cm]{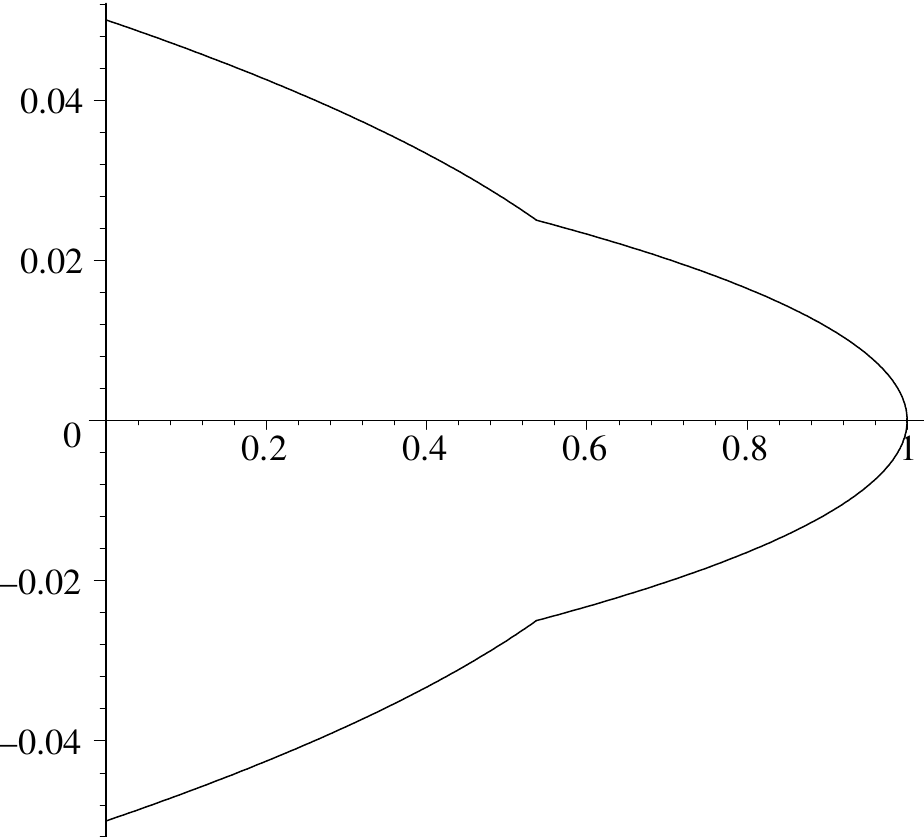}\;
		\includegraphics[width=5cm]{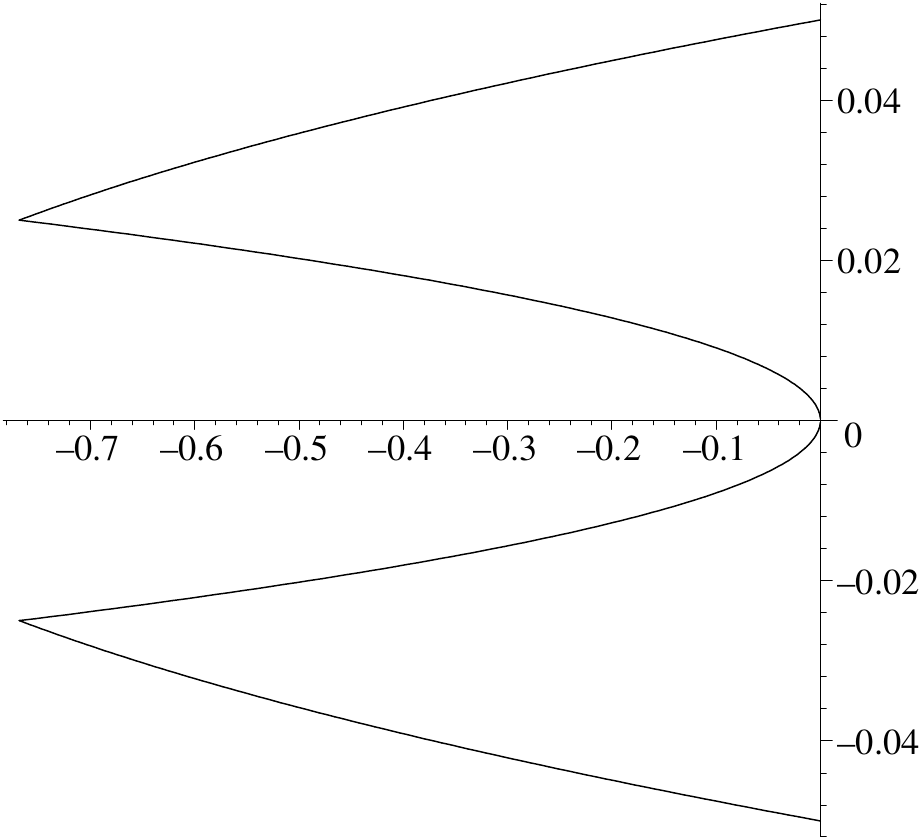}\;
		\includegraphics[width=5cm]{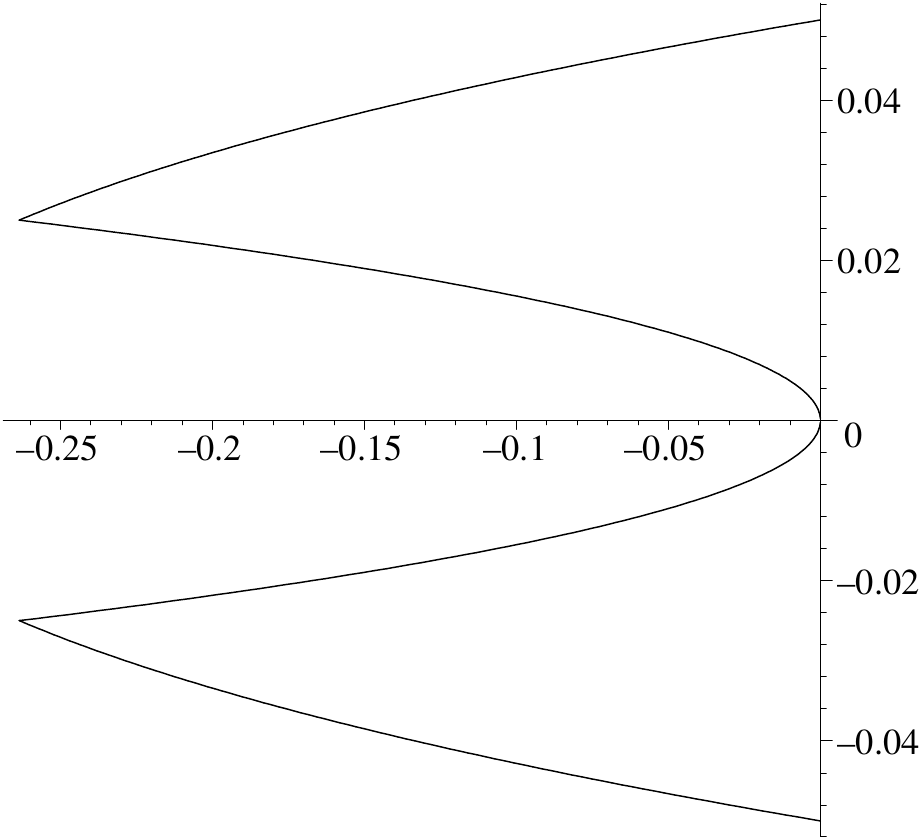}
	\caption{Warping functions and their derivatives for the case of a $[30/-30]_s$ laminate.}
	\label{fig:Orthotropic_p30m30m30p30}
\end{sidewaysfigure}

\textbf{The $[30/-30]_s$ laminate:} The same remark as above applies. Results are presented in figure~\ref{fig:Orthotropic_p30m30m30p30}.
\section{Conclusion}
Transverse shear warping functions for an equivalent single layer plate model have been built from a variational approach. Based on a plate model which has been previously used by other authors, the part of the strain energy which involves the shear phenomenon is expressed in function of the warping functions and their derivatives. Then the variational calculus leads to a differential system of equations which warping functions must verify. 
\par
However, solving this system requires the choice of values for the (global) shear strains and their derivatives. This is a crucial point because the chosen values may change considerably the shapes of the warping functions. For cross-ply laminates, a particular choice, whose justification is given, lead to excellent results in terms of the warping functions. For angle ply laminates, the above choice cannot be justified, but at this time, no other choice has been found relevant.
\par
For single layer isotropic and orthotropic plates, an analytical expression of the warping functions is given. They involve hyperbolic trigonometric functions. They differ from the $z-4/3z^3$ Reddy's formula which has been found to be a limit of present warping functions for isotropic and moderately thick plates. When the $h/L$ and/or the $G_{13}/E_{1}$ ratios significantly differ from those of isotropic and moderately thick plates, a difference between present warping functions and Reddy's formula can be observed, even for the isotropic single layer plate. Finite element simulations agree perfectly with the present warping functions in these cases.
\par
For multilayer cross-ply configurations, the warping functions are expressed from the differential system in the form of layer-wise analytical solutions involving matrix exponentials and integration constants which must be determined. Warping functions and their derivatives must verify conditions at the bottom, the middle, and the top of the plate and at each interface. These conditions form a last linear system whose unknowns are the expected constants. Present warping functions have been compared to results of 3D finite element simulations. They are in excellent agreement. 
\par
For angle-ply laminates, the above process gives warping functions that seem to have relevant shapes, even if the choice for global shear values cannot be justified, as it was told before. In addition, no finite element comparison has been presented because it is difficult to propose boundary conditions and prescribed load that permit to isolate the shear phenomenon. Further work must be done to clarify this point.   

\bibliographystyle{unsrt}
\bibliography{WFforAMLP}
\appendix
\label{dernierepage}
\end{document}